\documentclass[prd,twocolumn,nofootinbib]{revtex4-1}
\usepackage{lipsum} % placeholder text
\usepackage{multirow,xspace}
\usepackage{amsmath,amsfonts,amssymb}
\usepackage{graphicx}
\usepackage{ulem,fancyvrb}
\usepackage{hyperref}

\usepackage{cleveref}
\crefname{section}{Section}{Sections}
\crefname{equation}{Eq.}{Eqs.}
\crefname{figure}{Fig.}{Figs.}
\crefname{table}{Table}{Tables}

\usepackage[usenames,dvipsnames,svgnames,table]{xcolor}
\usepackage{xfrac}

\let\oldsqrt\sqrt
\def\sqrt{\mathpalette\DHLhksqrt}
\def\DHLhksqrt#1#2{%
\setbox0=\hbox{$#1\oldsqrt{#2\,}$}\dimen0=\ht0
\advance\dimen0-0.2\ht0
\setbox2=\hbox{\vrule height\ht0 depth -\dimen0}%
{\box0\lower0.4pt\box2}}
\DeclareMathOperator{\pois}{Pois}
\DeclareMathOperator{\sgn}{sgn}

\DeclareMathOperator{\U}{U}
\DeclareMathOperator{\SU}{SU}
\DeclareMathOperator{\SO}{SO}

\newcommand{\mo}{\ensuremath{m_0}\xspace}
\newcommand{\mhf}{\ensuremath{\widetilde{m}_{1/2}}\xspace}
\newcommand{\az}{\ensuremath{A_0}\xspace}
\newcommand{\tb}{\ensuremath{\tan\beta}\xspace}
\newcommand{\sgnmu}{\ensuremath{\sgn\left(\mu\right)}\xspace}

\newcommand{\cm}{\ensuremath{\sqrt{s}}\xspace}

\newcommand{\TeV}{\ensuremath{\,\mathrm{Te\kern -0.08em V}}\xspace}
\newcommand{\GeV}{\ensuremath{\,\mathrm{Ge\kern -0.08em V}}\xspace}
\newcommand{\MeV}{\ensuremath{\,\mathrm{Me\kern -0.08em V}}\xspace}
\newcommand{\fb}{\ensuremath{\,\mathrm{fb}}\xspace}
\newcommand{\ifb}{\ensuremath{\,\mathrm{fb}^{-1}}\xspace}

\newcommand{\tev}{\ensuremath{\mathrm{Te\kern -0.08em V}}\xspace}
\newcommand{\gev}{\ensuremath{\mathrm{Ge\kern -0.08em V}}\xspace}

\newcommand{\hz}{\ensuremath{h^0}\xspace}

\newcommand{\g}{\ensuremath{\widetilde{g}}\xspace}

\newcommand{\na}{\ensuremath{\widetilde{\chi}_1^0}\xspace}
\newcommand{\nb}{\ensuremath{\widetilde{\chi}_2^0}\xspace}
\newcommand{\nc}{\ensuremath{\widetilde{\chi}_3^0}\xspace}
\newcommand{\nd}{\ensuremath{\widetilde{\chi}_4^0}\xspace}

\newcommand{\cha}{\ensuremath{\widetilde{\chi}_1^\pm}\xspace}

\newcommand{\chabar}{\ensuremath{\widetilde{\chi}_1^\mp}\xspace}

\newcommand{\smu}{\ensuremath{\widetilde{\mu}}\xspace}

\newcommand{\ser}{\ensuremath{\widetilde{e}_R}\xspace}

\newcommand{\smr}{\ensuremath{\widetilde{\mu}_R}\xspace}
\newcommand{\sta}{\ensuremath{\widetilde{\tau}_1}\xspace}
\newcommand{\stb}{\ensuremath{\widetilde{\tau}_2}\xspace}

\newcommand{\numu}{\ensuremath{\widetilde{\nu}_\mu}\xspace}

\newcommand{\stopa}{\ensuremath{\widetilde{t}_1}\xspace}
\newcommand{\stopb}{\ensuremath{\widetilde{t}_2}\xspace}
\newcommand{\ba}{\ensuremath{\widetilde{b}_1}\xspace}

\newcommand{\q}{\ensuremath{\widetilde{q}}\xspace}
\newcommand{\slep}{\ensuremath{\tilde{\ell}}\xspace}
\newcommand{\snu}{\ensuremath{\widetilde{\nu}}\xspace}
\newcommand{\neut}{\ensuremath{\widetilde{\chi}^0}\xspace}
\newcommand{\chx}{\ensuremath{\widetilde{\chi}^\pm}\xspace}
\newcommand{\stau}{\ensuremath{\widetilde{\tau}}\xspace}

\newcommand{\bsmumu}{\ensuremath{B^0_s\to \mu^+\mu^-}\xspace}
\newcommand{\btaunu}{\ensuremath{B^+\to \tau^+\nu}\xspace}
\newcommand{\bsg}{\ensuremath{B\to X_s\gamma}\xspace}
\newcommand{\br}[1]{\ensuremath{\mathcal{B}r\left(#1\right)}\xspace}

\newcommand{\order}[1]{\ensuremath{\mathcal{O}\left(#1\right)}\xspace}

\newcommand{\MSBar}[1]{\ensuremath{#1^{\overline{\mathrm{MS}}}}\xspace}
\newcommand{\MBMB}{\ensuremath{m_b(m_b)}\xspace}
\newcommand{\AlphaSMZ}{\ensuremath{\alpha_\mathrm{s}(M_Z)}\xspace}

\newcommand{\AlphaEMMZInv}{\ensuremath{\alpha_\mathrm{EM}^{-1}(M_Z)}\xspace}

\newcommand{\gsugra}{\ensuremath{{}_{\g}\text{SUGRA}}\xspace}

\newcommand{\ir}[1]{\ensuremath{\mathbf{#1}}\xspace}

\newcommand{\PSet}{\ensuremath{\mathbf{\Theta}}\xspace}
\newcommand{\DSet}{\ensuremath{\mathbf{D}}\xspace}
\newcommand{\Mass}[1]{#1\text{ Mass (\gev)}\xspace}

\newcommand{\mh}{m_{\hz}}
\newcommand{\mg}{m_{\g}}
\newcommand{\mstau}{m_{\stau}}

\newcommand{\mpl}{M_\mathrm{Pl}}

\newcommand{\NEUAffil}{Department of Physics, Northeastern University, Boston, MA 02115, USA}
\newcommand{\susykit}{{\sc SusyKit}\xspace}

\allowdisplaybreaks[1]

\begin{document}
\title{Gluino-driven Radiative Breaking, Higgs Boson Mass, Muon $\mathbf{g-2}$, and the Higgs Diphoton Decay in SUGRA Unification}
\author{Sujeet~Akula}
\email[Email: ]{s.akula@neu.edu}
\affiliation{\NEUAffil}

\author{Pran~Nath} 
\email[Email: ]{nath@neu.edu}
\affiliation{\NEUAffil}

\begin{abstract}
  We attempt to reconcile seemingly conflicting experimental results on the Higgs boson mass,   
  the   anomalous magnetic moment of the muon, null results in search for supersymmetry  at the LHC within the 8\TeV data and results from $B$-physics, all within the context of 
  supersymmetric grand unified theories.
Specifically, we consider a supergravity grand unification model   
    with non-universal gaugino masses where we take the $\mathrm{SU}(3)_C$ gaugino field to be 
  much heavier than the other gaugino and sfermion fields at the unification scale. This construction naturally leads to a large mass splitting 
  between the slepton and squark masses, due to the mass splitting between the electroweak gauginos and the gluino. 
  The heavy Higgs bosons and Higgsinos also follow the gluino toward large masses.
  We carry out a Bayesian Monte Carlo analysis of the parametric space
  and find that it can simultaneously explain the large Higgs mass, and the anomalous magnetic moment
  of the muon,   while producing a negligible correction to   
   the Standard Model prediction for \br\bsmumu.  We also find that the model leads to an excess in the Higgs diphoton decay rate.
  A brief discussion of the possibility of detection of the light particles is given. Also 
  discussed are the implications of the model for dark matter.
\end{abstract}

\keywords{Higgs, Muon Anomalous Magnetic Moment, B-physics, Dark Matter, LHC, Supersymmetry}

\maketitle

\section{Introduction \label{intro}}
The CMS and ATLAS collaborations have discovered and measured~\cite{CMS:2012ufa, ATLAS:2012tfa, CMS:2012nga, ATLAS:2012oga, CMS:2013lba} 
the mass of a new boson which is most likely the Higgs boson~\cite{Englert:1964et, Higgs:1964ia, Higgs:1964pj, Guralnik:1964eu} responsible for breaking electroweak symmetry. 
In 
supersymmetry, one would identify this as the light $CP$-even Higgs boson~\cite{Akula:2011aa,Baer:2011ab, Arbey:2011ab, Draper:2011aa, Carena:2011aa,Akula:2012kk, Strege:2012bt}, \hz. 
Both experiments agree that the mass is between 125 and 126\GeV. 
It is quite remarkable that the observed Higgs boson mass lies 
close to the upper limit predicted in  grand unified supergravity models~\cite{Chamseddine:1982jx,Nath:1983aw,Hall:1983iz,Arnowitt:1992aq} 
which is roughly 130\GeV~\cite{Akula:2011aa,Akula:2012kk,Arbey:2012dq,Ellis:2012aa,Buchmueller:2011ab,Baer:2012mv}.
(For a recent review of Higgs and supersymmetry see~\cite{Nath:2012nh}.) 
Because the mass of the \hz boson in supersymmetry~\cite{Nath:1983fp,Carena:2002es,Djouadi:2005gj} is less than 
that of the $Z$ boson at the tree level, a large loop correction is necessary to match the measured 
value. The dominant one-loop Higgs self energy correction arises from its coupling to the top supermultiplet 
so that 

\begin{equation}
  \Delta \mh^2\simeq  \frac{3m_t^4}{2\pi^2 v^2} \ln \frac{M_{\rm S}^2}{m_t^2} 
    + \frac{3 m_t^4}{2 \pi^2 v^2}  \left(\frac{X_t^2}{M_{\rm S}^2} - \frac{X_t^4}{12 M_{\rm S}^4}\right)~,
  \label{tloop}
\end{equation}
where \(v=246\GeV\), $M_\mathrm{S}$ is the average stop mass, \linebreak \( X_t = A_t-\mu\cot\beta \), $\mu$ is the Higgs mixing 
parameter and $A_t$ is the trilinear coupling (both at the electroweak scale), and \( \tan\beta = \langle H_2\rangle/\langle H_1\rangle \), 
where $H_2$ gives mass to the up quarks while $H_1$ gives mass to the down quarks and leptons. 
Since  $\Delta \mh^2$ has a logarithmic dependence of $M_\mathrm{S}$, a sizable  $\Delta \mh^2$ correction 
implies that the scale $M_\mathrm{S}$  is high, lying in the several \tev region.

A high SUSY scale is also suggested by the ATLAS and CMS collaborations. 
So far, the LHC has delivered 5.3\ifb and 23\ifb of integrated luminosity~\cite{lhcluminosity} at 7\TeV and 8\TeV respectively to 
both CMS and ATLAS. Analysis of large portions of this data in search of supersymmetry has only yielded 
null results, though it is important to note that the parametric exclusion limits provided are typically 
only on minimal or simplified models. Whenever one works with non-minimal models of supersymmetry, it is 
necessary to evaluate the signal efficiencies specific to one's model and determine the credible 
region. The null searches can be evaded obviously by just raising the masses of the superpartners, 
and thereby raising the scale of SUSY, 
but it can also be done by producing mass hierarchies and mass splittings that are atypical in minimal 
models.

The search for the rare decay \bsmumu also has important implications for supersymmetry. 
The LHCb collaboration has recently observed~\cite{LHCb:2012nna} this rare decay, determining the 
branching ratio \(\br\bsmumu = (3.2^{+1.5}_{-1.2})\times10^{-9} \), which is in excellent agreement with 
the Standard Model, and thus requires the supersymmetric contribution~\cite{Choudhury:1998ze, Babu:1999hn, Bobeth:2001sq} to this decay to be very small.
This contribution is mediated by the neutral Higgs bosons and will involve a flavor-changing scalar quark loop.
(It is also sensitive to $CP$ violation~\cite{Ibrahim:2002fx, Ibrahim:2007fb}.) 
In the large \tb limit, the branching ratio is approximately~\cite{Degrassi:2006eh,Akula:2011ke}
\begin{multline}
  \label{burasetal}
  \br\bsmumu \simeq 3.5\times10^{-5}
     \left(\frac{\tau_{B_s}}{1.5\,\mathrm{ps}}\right) \left(\frac{f_{B_s}}{230\MeV}\right)^2 \\
     \times \left(\frac{\left|V_{ts}^\mathrm{eff}\right|}{0.040}\right)^2
     \left(\frac{\tb}{50}\right)^6 \left(\frac{m_t}{m_A}\right)^4 \\
     \times \frac{(16\pi^2)^2\epsilon_Y^2}{(1+(\epsilon_0 + \epsilon_Y y^2_t)\tan\beta)^2(1+\epsilon_0\tan\beta)^2} ~,
\end{multline}
where $\tau_{B_s}$ is the mean lifetime, $f_{B_s}$ is the decay constant, and $V_{ts}^\mathrm{eff}$ is the 
effective CKM matrix element. The loop factors $\epsilon_0$ and $\epsilon_Y$ are given in terms of soft breaking parameters 
of the 3rd generation $m_{\widetilde Q}$, $m_{\widetilde U}$, $m_{\widetilde D}$, which are 
the masses of the left-handed squark, 
up-type squark, and down-type squark,
 as well as the  gluino mass $\mg$, the strong coupling constant $\alpha_s$,
and the $CP$-odd Higgs mass $m_A$:
\begin{gather}
  \epsilon_0 = -\frac{2\alpha_s}{3\pi} \frac{\mu}{\mg} H(m_{\widetilde Q}^2/\mg^2, m_{\widetilde D}^2/\mg^2)\\
  \epsilon_Y = \frac{1}{16\pi^2} \frac{A_t}{\mu} H(m_{\widetilde Q}^2/\mu^2, m_{\widetilde U}^2/\mu^2) \\
 \hspace{-3pt} H(x_1,x_2)=\frac{x_1 \ln x_1}{(1-x_1)(x_1-x_2)}+\frac{x_2 \ln x_2}{(1-x_2)(x_2-x_1)}.
\end{gather}

We note that the  branching ratio given by \cref{burasetal} 
is suppressed by the factor 
$(m_t/m_A)^4$ and so a large weak scale of SUSY which implies a large $m_A$, naturally leads to a 
small contribution to \br\bsmumu. Additionally, we see in \cref{burasetal} the factor
$(\tan\beta/50)^6$, which implies that the SUSY contribution to \bsmumu is further suppressed if $\tb\lesssim50$.
Together these effects also reduce the SUSY contribution~\cite{Bertolini:1990if,Degrassi:2006eh} to \br\bsg
to negligible value.

While the  observation of a high Higgs boson mass, null results on the discovery of sparticles and 
the observation of no significant deviation in the \bsmumu branching ratio from the Standard Model result all appear to 
indicate a high scale for SUSY, the opposite is indicated by the Brookhaven experiment E821~\cite{Bennett:2006fi} 
which measures
\(a_\mu = \frac{1}{2}(g_\mu-2)\) to deviate from the Standard Model prediction~\cite{Hagiwara:2011af, Davier:2010nc} at the $3\,\sigma$ level. If this deviation is taken 
to arise from supersymmetry, then 
\begin{equation}
  a_\mu^\mathrm{SUSY} = \delta a_\mu = (287 \pm 80.)\times10^{-11} ~.
  \label{bnl}
\end{equation}
The SUSY contribution~\cite{Yuan:1984ww, Kosower:1983yw,Lopez:1993vi,Chattopadhyay:1995ae,Moroi:1995yh,Ibrahim:1999aj,Heinemeyer:2003dq,Sirlin:2012mh}
arises from \chx--\numu and \neut--\smu loops.
A rough estimate of the supersymmetric correction is
\begin{equation}
  \delta a_{\mu} \simeq \sgn(M_2 \mu) \left(130 \times 10^{-11}\right) \left(\frac{100\GeV}{M_\mathrm{SUSY}}\right)^2 \tb ~,
\end{equation}
where $M_\mathrm{SUSY}$ is the SUSY scale.
In order to obtain a  SUSY correction of size indicated by \cref{bnl} the masses of sparticles in the loops,
i.e., the masses of 
 \chx, \neut, \smu, and \numu  must be only about a few hundred \gev.

Another result which may be a signal of SUSY concerns the  excess seen in the 
diphoton decay rate of the Higgs, which is above the Standard Model prediction. This excess is 
parametrized by the signal strength 
\begin{equation}
  R_{\gamma\gamma} = \frac{\sigma(pp\to H)_\mathrm{obs}}{\sigma(pp\to H)_\mathrm{SM}} \times \frac{\Gamma(H\to\gamma\gamma)_\mathrm{obs}}{\Gamma(H\to\gamma\gamma)_\mathrm{SM}}
  \label{rggdef}
\end{equation}
and is reported as $R_{\gamma\gamma} = 1.6\pm0.4$ at CMS~\cite{CMS:2012nga} and $R_{\gamma\gamma} = 1.8\pm0.5$ at 
ATLAS~\cite{ATLAS:2012oga}.
The excess is not statistically conclusive and can easily be attributed to a simple fluctuation
or to QCD uncertainties~\cite{Baglio:2012et}.
Still it is worthwhile to consider how SUSY can contribute to this loop-induced decay (considering \hz in place of $H$). 
The excess in the diphoton rate has been discussed in a variety of models by various authors
(see, e.g., 
~\cite{Carena:2011aa, Giudice:2012pf, Feng:2013mea} and the references  therein).
Within the MSSM, the largest contributions would arise via a \stau triangle, provided that its mass is not too high. 
(We discuss the calculation of $R_{\gamma\gamma}$ in more detail in \cref{diphoton}.) 
So, if the diphoton result is real, we have another indication of low scale SUSY.

Assuming that the $g_{\mu}-2$ and the diphoton rate hold up, one has  apparently conflicting 
results for the weak scale of SUSY.  On the one hand, the high Higgs boson mass, 
 null results on the observation of sparticles at the LHC, and the lack of any significant deviation in the \br\bsmumu branching ratio from the Standard Model prediction 
point to a high SUSY scale, i.e., a SUSY scale lying in the several \tev range. On the other hand, 
 the $3\,\sigma$ deviation in $a_\mu$ and a fledgling excess in the diphoton decay of the Higgs boson decay 
point to a low SUSY scale lying in the sub-\tev range.  These results taken together, point to 
a split scale SUSY with one scale governing the colored sparticle masses and the heavy Higgs boson masses, and
the other SUSY scale governing the uncolored sparticle masses. To generate this split scale SUSY, we 
construct in this work a supergravity grand unified model~\cite{Chamseddine:1982jx, Nath:1983aw, Hall:1983iz} 
by introducing non-universalities in the gaugino sector with the feature that the gaugino mass in the
$SU(3)_C$ sector is much larger than the other soft masses. In this model, radiative electroweak symmetry breaking~\cite{Inoue:1982pi,Ibanez:1982fr,AlvarezGaume:1983gj} 
(for a review see~\cite{Ibanez:2007pf}) 
 is driven by the gluino mass. In this work, we label this model as \gsugra. 
We will show that \gsugra satisfies 
all of the experimental results simultaneously by exploiting a feature of the renormalization group equations which leads 
to a splitting between the squarks, gluino, Higgs bosons, and Higgsinos which become very heavy, and the sleptons, 
bino and winos which are allowed to remain light at the electroweak scale. (The sfermion masses still unify at a 
high scale.) We will use a Bayesian Monte Carlo analysis of \gsugra to show that it satisfies all experimental 
results and determine the credible regions in the parameters and sparticle masses.

The  outline  of the rest of the paper is as follows:  In \cref{model}, we discuss
the general framework of non-universal SUGRA models with specific focus on \gsugra where the gaugino
mass in the $SU(3)_C$ color sector is much larger than other mass scales in the model. 
In \cref{framework}, we discuss  the statistical framework used 
in our Bayesian Monte Carlo analysis of a simplified parametric space for \gsugra. In \cref{lhc} we explore the impact of LHC 
searches for sparticles on \gsugra using event-level data and signal simulations. The results of our analyses 
as well as the details of Higgs diphoton rate are presented 
in \cref{results}. Concluding remarks are given in \cref{conclusion}.

\section{\boldmath The \gsugra Model \label{model}}

Supergravity grand unification~\cite{Chamseddine:1982jx, Nath:1983aw, Hall:1983iz} is a broad framework 
which depends on three arbitrary functions: the superpotential, the K\"ahler potential, and the gauge kinetic 
energy function. Simplifying assumptions on the K\"ahler potential and the gauge kinetic energy function lead to 
universal boundary conditions for the soft parameters which is the basis of the model referred to as mSUGRA/CMSSM. 
The parameter space of mSUGRA is given by \mo, $m_{1/2}$, \az, \tb, and \sgnmu, where \mo is the universal scalar mass,
$m_{1/2}$ is the universal gaugino mass, \az is the universal trilinear coupling, and \( \tan\beta = \langle H_2\rangle/\langle H_1\rangle \). 
Here $H_2$ gives mass to the up quarks and $H_1$ gives mass to the down quarks and the leptons,
and $\mu$ is the Higgs mixing parameter which enters in the superpotential as $\mu H_1H_2$.

However, the supergravity grand unification framework does allow for non-universalities of the soft parameters, i.e., non-universalities for the scalar masses,
for the trilinear couplings and for the gaugino masses\footnote{The literature on non-universalities in SUGRA models is enormous. For a sample of early and later works
see~\cite{Ellis:1985jn,Drees:1985bx,Nath:1997qm,Ellis:2002wv,Anderson:1999uia,Huitu:1999vx,Corsetti:2000yq,
Chattopadhyay:2001mj,Chattopadhyay:2001va,Martin:2009ad,Feldman:2009zc,Gogoladze:2012yf,Ajaib:2013zha}
 and for a review see~\cite{Nath:2010zj}.}.
In \gsugra, we consider supergravity grand unification with universal boundary conditions in all sectors  except in the 
gaugino sector. In this sector, we specify that the $\SU(3)_C$ gaugino mass, $M_3$, be much larger than the universal
scalar mass and also much larger than the gaugino masses $M_2, M_1$ in the $\SU(2)_L$, $\U(1)_Y$ sectors, i.e., 
\begin{equation}
  M_3 \gg m_0, M_1, M_2
  \label{1.2}
\end{equation}
The constraints of \cref{1.2}
 ensure that the radiative breaking of electroweak symmetry will be driven by the gluino (hence, \gsugra). Now, the gluino mass enters
in the renormalization group equations for the squark masses and thus the squark masses will be driven to values proportional to the
gluino mass as we move down from the GUT scale toward the electroweak scale. Consequently, a gluino mass in the ten \tev
region will also generate a squark mass in the several \tev region. On the other hand, the RGEs for the
sleptons do not depend on the gluino mass at the one-loop level and if $\mo, M_1, M_2$ are $\order{100\GeV}$, 
 the masses of the sleptons as well as the electroweak gauginos at the electroweak scale will likely remain this size. Thus the
RG evolution creates a natural splitting of masses between the squarks and the sleptons at the  
electroweak scale even though they have a common mass at the grand unification scale. The renormalization of these 
soft masses for a sample point in \gsugra is shown in \cref{RGE}.
The huge mass splitting between the squark and slepton masses at low scales even though they are unified at 
high scales 
 is reminiscent of the gauge coupling unification where the three gauge couplings
$\alpha_i$ which are split at the electroweak scale but come together 
at the grand unification scale. We note that the split spectrum of \gsugra is very different in nature from that of what is commonly called 
``split supersymmetry''~\cite{ArkaniHamed:2004fb}, which consists of light Higgsinos $\widetilde H_{u,d}$, $\widetilde B$, $\widetilde W$, \g and one 
Higgs doublet but does not allow for light sfermions. 

In GUT models, non-universal gaugino masses can arise from superfields that transform as a non-singlet IRs of the GUT group and 
get VEVs in the spontaneous breaking and give masses to the gauginos. The general form of the gaugino mass term in the Lagrangian is 
\begin{equation}
  -\frac{\langle F\rangle_{ab}}{\mpl} \frac{1}{2} \lambda_a \lambda_b + {\rm H.c.} 
  \label{gmass1}
\end{equation}
where $\langle F\rangle_{ab}$ is a non-zero VEV of mass dimension 2, and $\mpl$ is the Planck mass. The
 $\lambda$'s belong to the adjoint of the 
GUT group: \ir{24} for $\SU(5)$ and \ir{45} for $\SO(10)$.  Now only the symmetric product
of the adjoints enters in the analysis. Thus for $\SU(5)$ one has 
$(\ir{24}\otimes \ir{24})_\mathrm{sym} = \ir{1} \oplus \ir{24} \oplus \ir{75} \oplus \ir{200}$, 
 while for $\SO(10)$ one has 
$(\ir{45}\otimes \ir{45})_\mathrm{sym} =  \ir{1} \oplus \ir{54} \oplus \ir{210} \oplus \ir{770}$.
With the use of singlet and non-singlet breaking, one can produce a hierarchy in the gaugino masses 
so that \cref{1.2} holds. 
We note that non-universalities of gaugino masses arise also in string based models, see, e.g., \cite{Kaufman:2013pya}.

In our study of \gsugra, we introduce gaugino sector non-universalities by having $m_{1/2}\to\mhf\equiv M_1=M_2$ 
and $M_3=10\,\mhf$ as an illustrative example, so that at the unification scale,  \(M_1:M_2:M_3 = 1:1:10\). We 
now show how this choice can be constructed by combining singlet and non-singlet breaking in $\SU(5)$ and
in $\SO(10)$.  In $\SU(5)$ we consider the linear combination \(\ir{1} + a\,\ir{24} + b\,\ir{75}\). 
Now the  singlet breaking gives the ratio $M_1:M_2:M_3= 1:1:1$, the 24-plet gives the ratio~\cite{Martin:2009ad}
$(-1/2:-3/2:1)$ while the 75-plet gives the ratio~\cite{Martin:2009ad}  $(-5:3:1)$. Choosing $a= -8/11$ and $b=-1/11$ 
leads to the desired ratio $M_1:M_2:M_3= 1:1:10$. This scheme also applies to $\SO(10)$ 
since $\SU(5) \subset \SO(10)$. However, for $\SO(10)$
we can also consider gaugino mass terms in representations of
$\SU(4)\times \SU(2)_L \times \SU(2)_R \subset \SO(10)$ and label the breaking terms  by 
$\SU(4)\times \SU(2)_R$ representations as subscripts. In this case we consider the breaking 
$\ir1+ a\,\ir{210}_{(\ir1,\ir1)} + b\,\ir{210}_{(\ir{15},\ir1)}$ where the $\ir{210}_{(\ir1,\ir1)}$ gives the gaugino 
mass ratio~\cite{Martin:2009ad} 
of $(-3/5:1:0)$ and $\ir{210}_{(\ir{15},\ir1)}$ gives the gaugino mass ratio~\cite{Martin:2009ad} of $(-4/5:0:1)$.  Thus we can choose 
$a=-3/4$ and $b=3/2$ to get the desired $1:1:10$ ratio. We limit ourselves to this ratio for the rest of the
analysis in this paper. 
However, many features of this analysis will persist with different ratios of $M_1:M_2:M_3$ as long as $M_3 \gg \mo, M_1, M_2$. 

In \gsugra, radiative electroweak
symmetry breaking is dominated by the large gluino mass which is responsible for giving large masses
to the squarks.  We contrast this work with other recent works which have attempted to explain $g_{\mu}-2$ in the
context of a high Higgs boson mass.  This is attempted in~\cite{Giudice:2012pf} with the assumption of a light 
slepton and heavy squark spectrum. The analysis also tries to correlate $g_{\mu}-2$ with the diphoton 
rate. However, this model is not a high scale model and the analysis is limited to assumptions of the spectrum
at the electroweak scale. In~\cite{Ibe:2013oha} the authors assumed a split family supersymmetry.
The analysis of~\cite{Mohanty:2013soa} uses non-universal gaugino masses in an $\SU(5)$ model 
but the details of the model are significantly different from the work presented here. The work
\cite{Bhattacharyya:2013xba} also addresses the issue of getting light uncolored and heavy colored 
particles but the analysis is within a gauge mediated supersymmetry breaking.

\begin{figure}[t!]
  \begin{center}
    \includegraphics[scale=0.30]{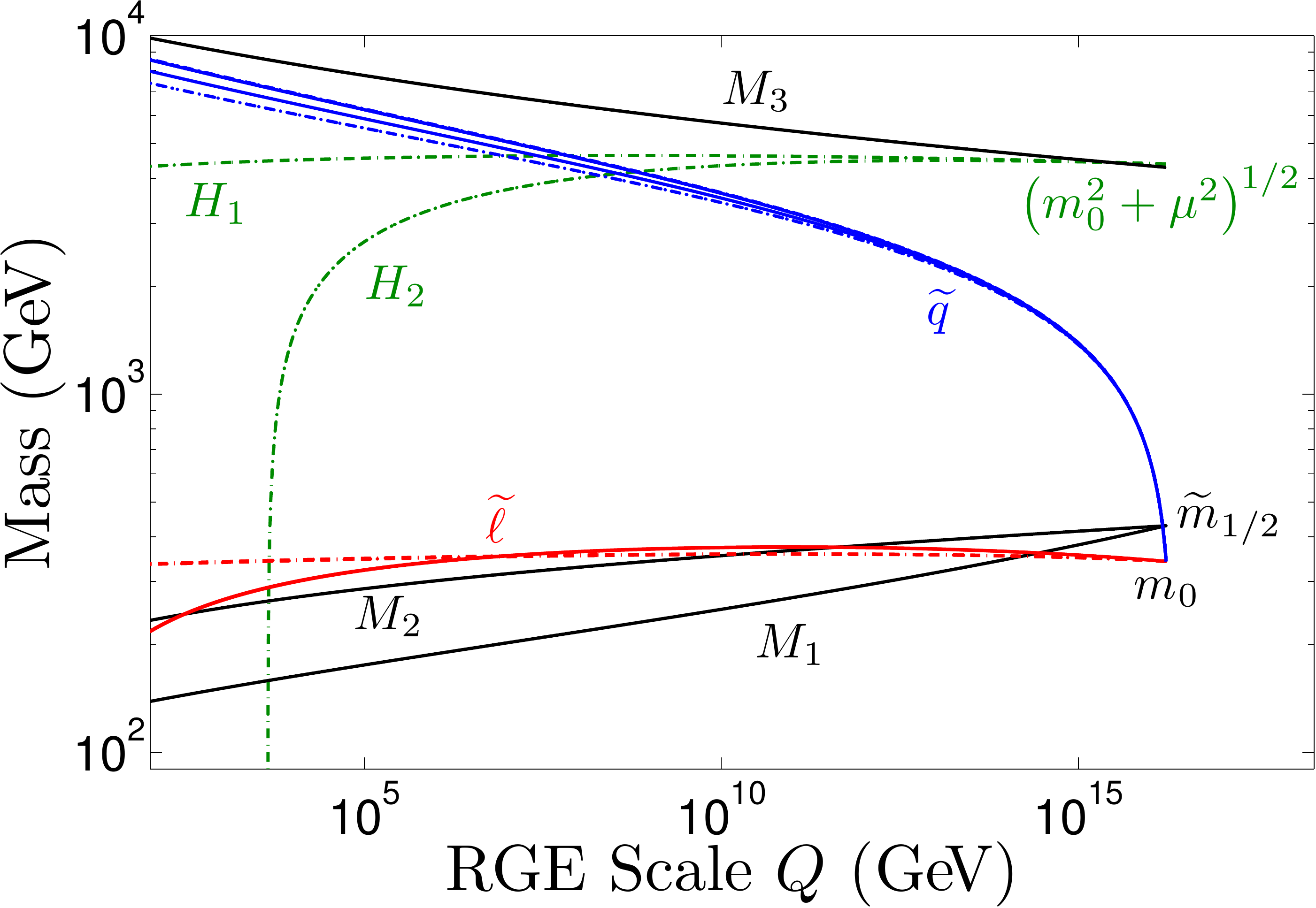}
    \caption{\label{RGE}
      Two-loop renormalization group evolution of the soft parameters in \gsugra. The input parameters used here are those of the 
      best-fit point determined from our analysis in \cref{framework}. The fields are labeled in the figure and also in color. The gaugino fields are presented in 
      black and the Higgs fields are presented in green. The squarks and sleptons are in blue and red, 
      where the left-handed fields are solid and the right-handed fields are dot-dashed. Additionally, \mo is the soft  mass 
      for the scalars, \mhf is the common mass of the $\U(1)_Y$ and $\SU(2)_L$ gaugino fields, and $\mu$ is the Higgs 
      mixing parameter. 
    }
  \end{center}
\end{figure}

The attractive feature of \gsugra is that the relatively large value of $M_3$ automatically drives the squarks
to be massive while the sleptons as well as the bino and the light wino are left alone. This is illustrated in 
\cref{RGE} where we display the renormalization group flow for a sample point from our analysis. We wish to show that 
this simple feature automatically satisfies all of the empirical results that we have discussed here that 
hint at the supersymmetric spectrum. To this end, we perform a Bayesian Monte Carlo analysis of \gsugra with the 
illustrative example of the $(1:1:10)$ gaugino mass ratio, which we discuss in the sections that follow.

\begin{table*}[t!]
  \begin{center}
    \begin{tabular*}{0.9\textwidth}{@{\extracolsep{\fill}}| c | c c c | c | c |} \hline
      Observable		& Central value	& Exp. Error	& Th. Error	& Distribution		& Ref. \\ \hline\hline
      \multicolumn{6}{|c|}{SM Nuisance Parameters} \\ \hline\hline
      $m_t^\mathrm{pole}$ (\gev) 	&  $173.5$	& $1.0$	& --		& Gaussian 		& \cite{pdg2012}\\
      \MSBar\MBMB (\gev)	& $4.18$	& $0.03$	& --		& Gaussian		& \cite{pdg2012} \\
      \MSBar\AlphaSMZ 		& 0.1184 	& $7\times10^{-4}$ & --		& Gaussian		& \cite{pdg2012} \\
      \MSBar\AlphaEMMZInv	& $127.933$ 	& 0.014 	& --		& Gaussian	 	& \cite{pdg2012}\\ \hline\hline
      \multicolumn{6}{|c|}{Measured} 	\\ \hline\hline
      $\delta a_\mu\times10^{11}$ & 287		& 80		& 10		& Gaussian		& \cite{Bennett:2006fi, Hagiwara:2011af, Davier:2010nc}	\\
      $\br\bsmumu\times10^{9}$	& 3.2		& 1.92		& 14\%		& Gaussian 		& \cite{LHCb:2012nna}\\
      $\br\bsg\times10^{4}$ 	& 3.55		& 0.26		& 0.21		& Gaussian		& \cite{hfag2012}\\
      $\br\btaunu\times10^{4}$ 	& 1.79		& 0.48		& 0.38		& Gaussian		& \cite{hfag2012}\\
      $\omega_\chi$		& 0.1126	& 0.0036	& 10\%		& Upper-Gaussian	& \cite{WMAP:2010fb} \\
      \Mass\hz			& $125.7$	& $0.2$		& $2.0$		& Gaussian 		& \cite{CMS:2012nga, ATLAS:2012oga}\\ \hline\hline
      \multicolumn{6}{|c|}{95\%~CL Particle Mass Limits (\gev)}	\\\hline\hline
      \hz			& $122.5$	& --		& --		& Lower -- Step Func.	& \cite{CMS:2012tx} \\
      \hz			& $129$		& --		& --		& Upper -- Step Func.	& \cite{CMS:2012tx}\\
      \na			& $46$		& --		& 5\%		& Lower -- Error Func.	& \cite{pdg2012} \\
      \nb			& $62.4$	& --		& 5\%		& Lower -- Error Func.	& \cite{pdg2012} \\
      \nc			& $99.9$	& --		& 5\%		& Lower -- Error Func.	& \cite{pdg2012}\\
      \nd			& $116$		& --		& 5\%		& Lower -- Error Func.	& \cite{pdg2012}\\
      \cha			& $94$ 		& --		& 5\%		& Lower -- Error Func.	& \cite{pdg2012}\\
      \ser			& $107$		& --		& 5\%		& Lower -- Error Func.	& \cite{pdg2012} \\
      \smr			& $94$		& --		& 5\%		& Lower -- Error Func.	& \cite{pdg2012} \\
      \sta			& $81.9$	& --		& 5\%		& Lower -- Error Func.	& \cite{pdg2012}\\
      \ba			& $89$		& --		& 5\%		& Lower -- Error Func.	& \cite{pdg2012}\\
      \stopa			& $95.7$	& --		& 5\%		& Lower -- Error Func.	& \cite{pdg2012} \\
      \g			& $500$		& --		& 5\%		& Lower -- Error Func.	& \cite{pdg2012}\\
      \q			& $1100$	& --		& 5\%		& Lower -- Error Func.	& \cite{pdg2012} \\
      \hline
    \end{tabular*}
    \caption{\label{liketable} 
      Summary of the observables used to construct the likelihood function. The distribution labeled 
      ``Upper-Gaussian'' used for the $\omega_\chi$ observable means that there is only a 
      decrease in likelihood for values larger than the central value. The 95\%~CL limits are evaluated 
      using the complementary error function, as the bound is smeared by the theoretical uncertainty. Limits 
      specified with a step function distribution indicate a hard cut, where points on the wrong side of the 
      limit are assigned zero likelihood.
    }
  \end{center}
\end{table*}

\section{Statistical Framework \label{framework}}
We study here the parameter space of \gsugra for the case where the ratio of the gaugino masses at the GUT scale is $1:1:10$. 
In this case, \gsugra is parametrized by \mo, \mhf, \az, and \tb (having selected $\sgnmu=1$). Here, \(\mhf=M_1=M_2\) while \(M_3=10\,\mhf\). 
 The dimensionful parameters \mo, \mhf, and \az 
are all specified at the GUT scale. The ratio of the two Higgs VEVs $\tb = \langle H_2\rangle/\langle H_1\rangle$, 
is specified at $M_Z$.  
We further include four Standard Model nuisance parameters to create an 8D parameter space. Namely, we add the top quark pole 
mass, the running bottom quark mass, the strong coupling, and the EM coupling. 
We create from these the parameter space \PSet:
\begin{equation}
  \begin{split}
    \PSet = &\left\{\mo, \mhf, \az, \tb, m_t^\mathrm{pole}\hspace{-6pt}, \hspace{36pt}\right. \\
      & \hspace{24pt}\left. \MSBar\MBMB, \MSBar\AlphaSMZ, \MSBar\AlphaEMMZInv\right\} ~.
  \end{split} \label{paramset}
\end{equation}
For each parameter $\theta_i \in \PSet$, we begin by selecting uniform distributions in the allowed ranges prior to 
considering the experimental data. The prior distributions that we have selected for our parameters are uniform on 
either a linear or a log scale: 
\begin{equation}
  \begin{split}
    \mo \in [50, 5000]\GeV \text{ (log)} \\
    \mhf \in [50, 2500]\GeV \text{ (log)} \\
    \az \in [-50, 50]\TeV \text{ (linear)} \\
    \tb \in [3, 60] \text{ (linear)} ~.
  \end{split}  \label{priors}
\end{equation}
The nuisance parameters in \PSet are uniform in a $2\,\sigma$ range (linear scale) around the central values, which 
are specified in \cref{liketable}.

\begin{figure*}[t]
  \begin{center}
    \includegraphics[scale=0.65]{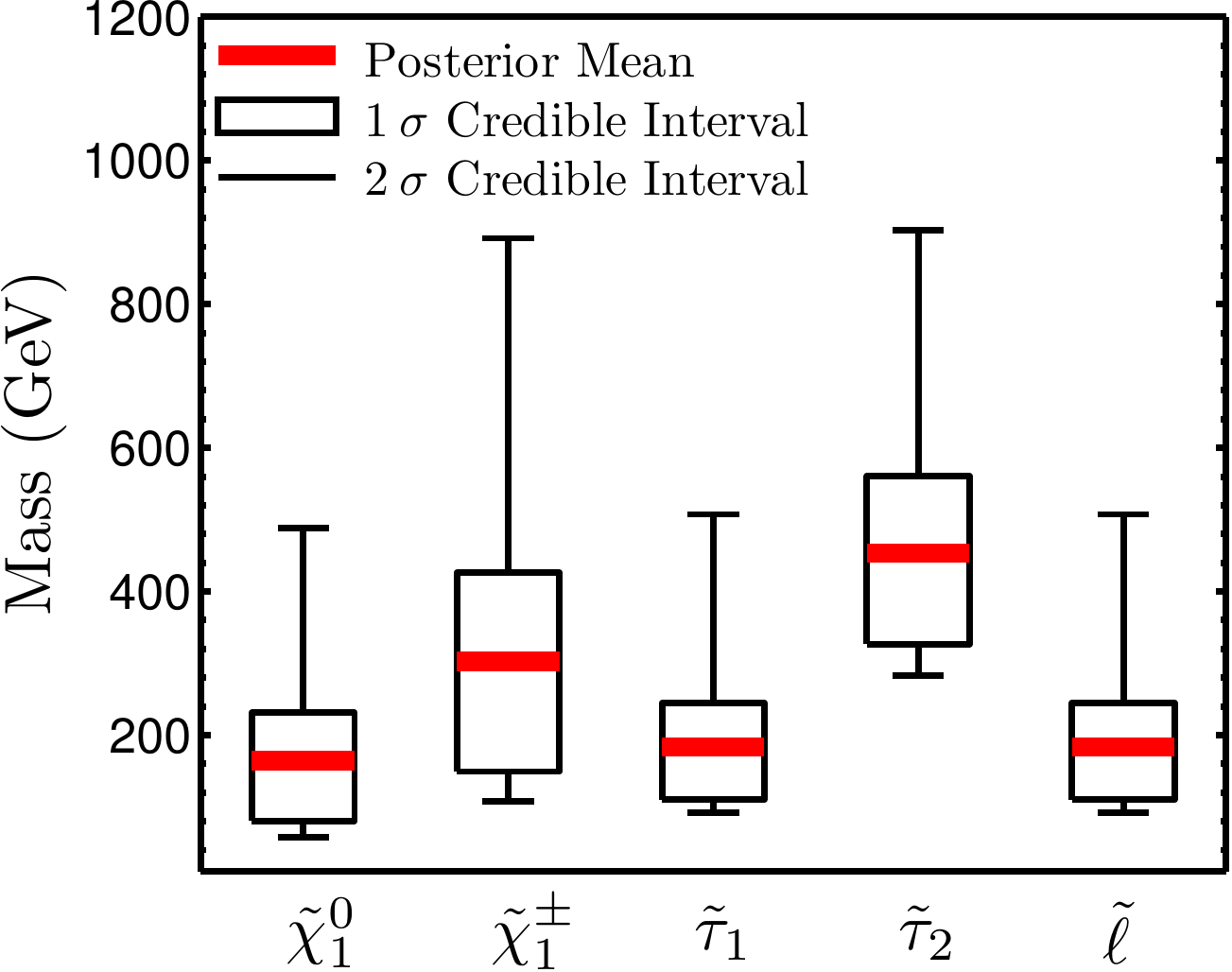}
    \hspace{0.5cm}
    \includegraphics[scale=0.65]{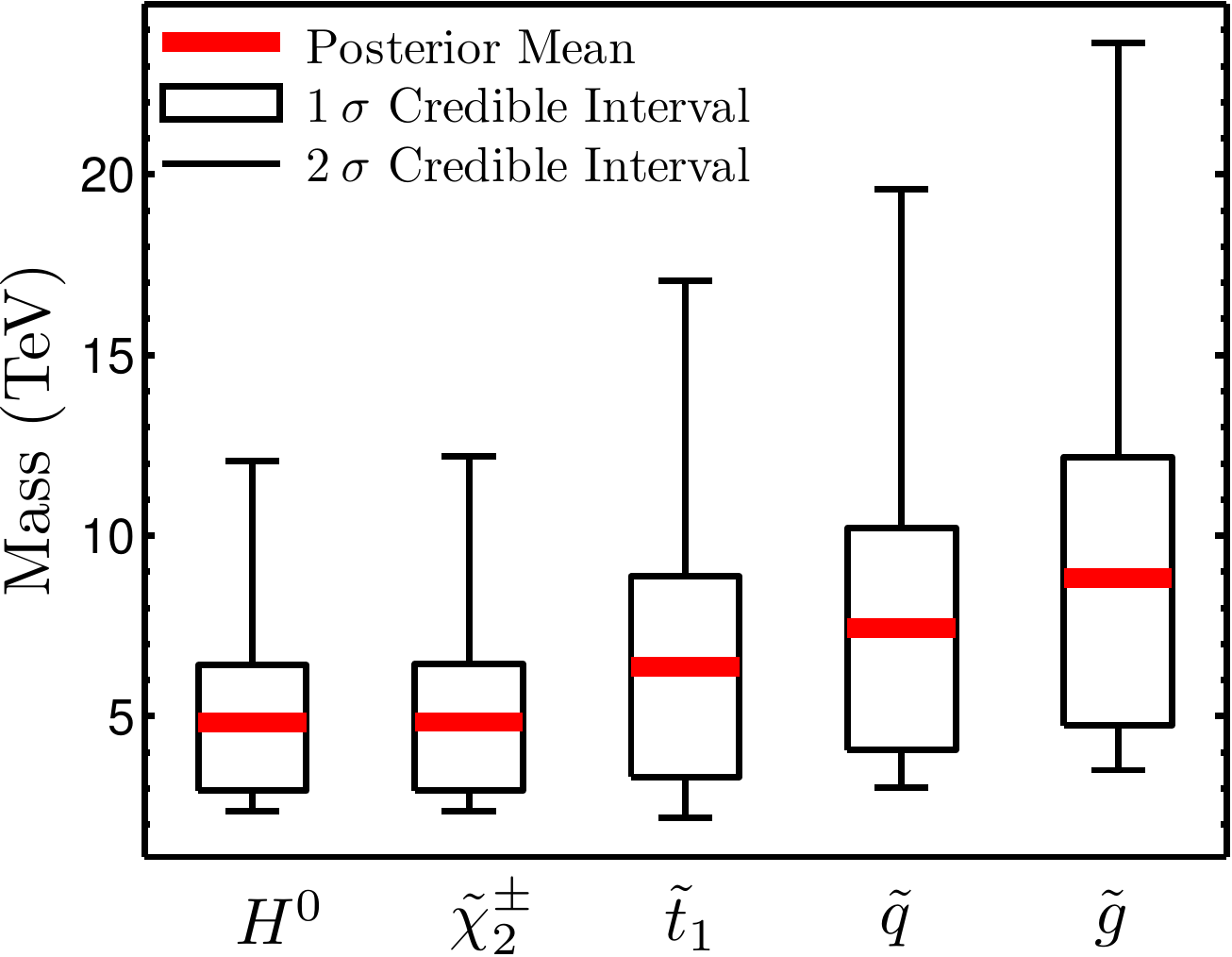}
    \caption{\label{masspost}
      A display of the mass spectrum  for sparticles and the Higgs boson with split scales, i.e., a low scale for $\tilde \chi_1^0, \tilde \chi^{\pm}_1, \tilde \tau_1,
      \tilde \tau_2, \tilde {\it l}$  and a high scale for $H^0, \tilde \chi^{\pm}_2, \tilde t_1, \tilde q, \tilde g$. 
         Shown are the credible intervals in the superpartner masses from the Bayesian analysis of \gsugra. The 
      lighter superpartners are presented in the 
      left panel, and the heavier are presented in the right panel. The posterior means are indicated in red.
    }
  \end{center}
\end{figure*}

Next we collect the relevant observables into \DSet, which is a set of pairs of central values and uncertainties 
of experimental measurements. The observables include the precise measurements of the nuisance parameters, along 
with the results from flavor physics \br\bsmumu and \br\bsg, the muon anomalous magnetic moment $\delta a_\mu$, 
the measured mass of the (ostensibly) light $CP$-even Higgs boson, as well as limits on superpartner masses. We 
further include the fit to the thermal relic density of dark matter, $\omega_\chi\equiv\Omega_\chi h^2$, from CMB temperature fluctuations 
measured by WMAP (9~year dataset)~\cite{WMAP:2012fq} and Planck (15.5~month dataset)~\cite{Planck:2013xsa}. In \gsugra, the lightest neutralino is indeed a 
candidate for cold dark matter, but we wish to allow for multicomponent models of dark matter, and so we only consider 
the upper limit of $\omega_\chi$. The central values and uncertainties of \DSet are specified in \cref{liketable}.

The goal now is to update our a priori guess for the probability distributions of the parameters in \PSet (given in \cref{priors}) with 
the empirical information in \DSet, giving the posterior probability distribution. This distribution can then be 
marginalized to determine the credible region of one or two parameters. The calculation of the posterior probability 
distribution is achieved using Bayesian inference, but we first need to be able to compare a parametric point in our 
model to the empirical data in \DSet. This requires a set of mappings $\xi_i:\PSet\to\mathbb{R}$ corresponding to each $d_i\in\DSet$, 
which just give the theoretical calculation for the observable corresponding to each $d_i$. These mappings are 
computed using numerical codes incorporated in our analysis software \susykit~\cite{susykit}.

Now we can move on to constructing the posterior probability distribution, which is given by \linebreak Bayes' theorem
\begin{equation}
  P(\PSet|\DSet) = \frac{P(\DSet|\PSet)P(\PSet)}{P(\DSet)} ~.
  \label{bayes}
\end{equation}
$P(\PSet)$ is the prior distribution given in \cref{priors}. The denominator is the so-called Bayesian evidence 
$\mathcal{Z} = P(D)$, which can be used in model selection tests, but as we are only interested in parameter 
estimation, it serves as a normalization constant. The final factor is the likelihood function 
$\mathfrak{L} = P(\DSet|\PSet)$, which is constructed by the ``pulls'' method
\begin{equation}
  -2\ln\mathfrak{L} = \sum\limits_{d_i\in\DSet} \frac{\left(\xi_i(\PSet) - d_i\right)^2}{\sigma_i^2 + \tau_i^2}
  \label{pulls}
\end{equation}
where $\sigma_i$ and $\tau_i$ are the experimental and theoretical uncertainties, respectively. This is straightforward 
for the case that a measurement with precision is reported. In many cases only the 95\%\,CL limits are given. In 
those cases, a smearing due to the implicit theoretical uncertainty in the computation is used and the likelihood is 
computed from the complementary error function. A hard cut on an observable can also be made by using a step function, 
i.e. assigning zero likelihood to points that are on the wrong side of a limit. The numerical values used to 
construct the likelihood function is given in \cref{liketable}.

\begin{figure*}[t!]
  \begin{center}
    \includegraphics[scale=0.30]{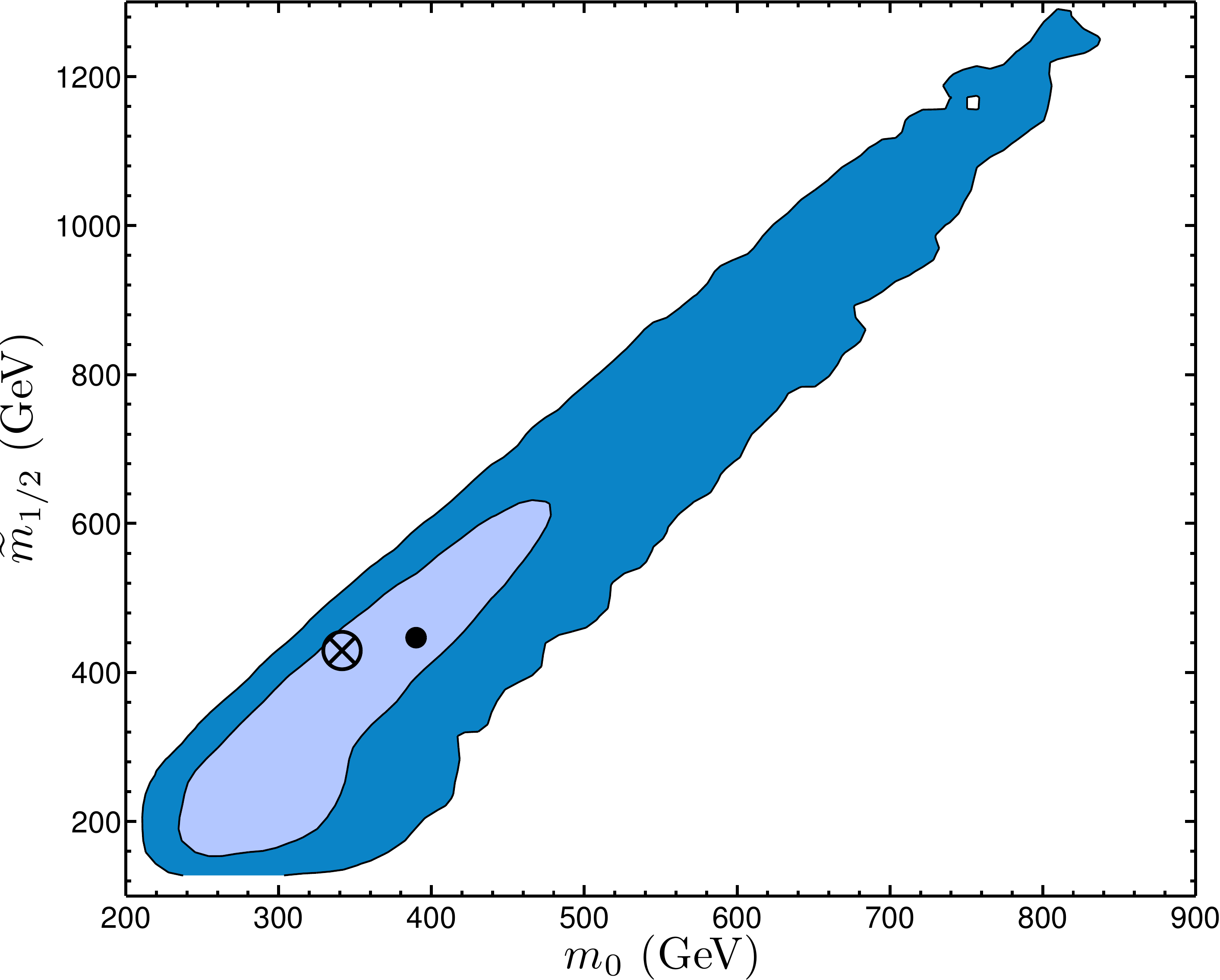}
    \includegraphics[scale=0.30]{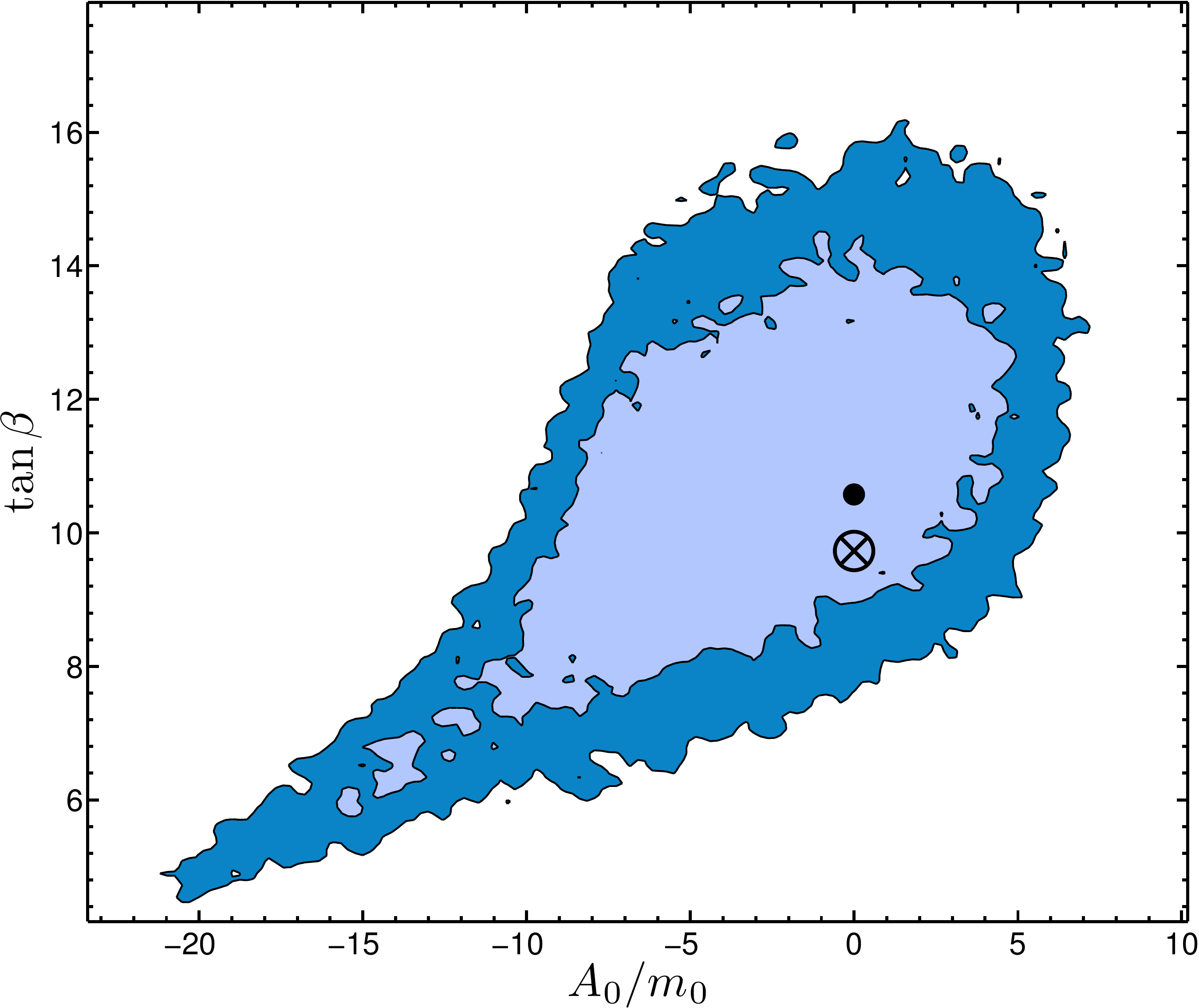}
    \caption{\label{twodplots}
      An exhibition of the $1\,\sigma$ and $2\,\sigma$ credible regions of the marginalized posterior probability distributions for 
      the parameters of interest of \gsugra. Left panel: the credible regions in \mo and \mhf. Right panel: the credible regions in 
      the dimensionless parameter $\az/\mo$ and \tb. The location of the best-fit point is  indicated by a circled `X' and the 
      posterior mean is given with a solid dot.
    }
  \end{center}
\end{figure*}
\begin{figure*}[t!]
  \begin{center}
    \includegraphics[scale=0.14]{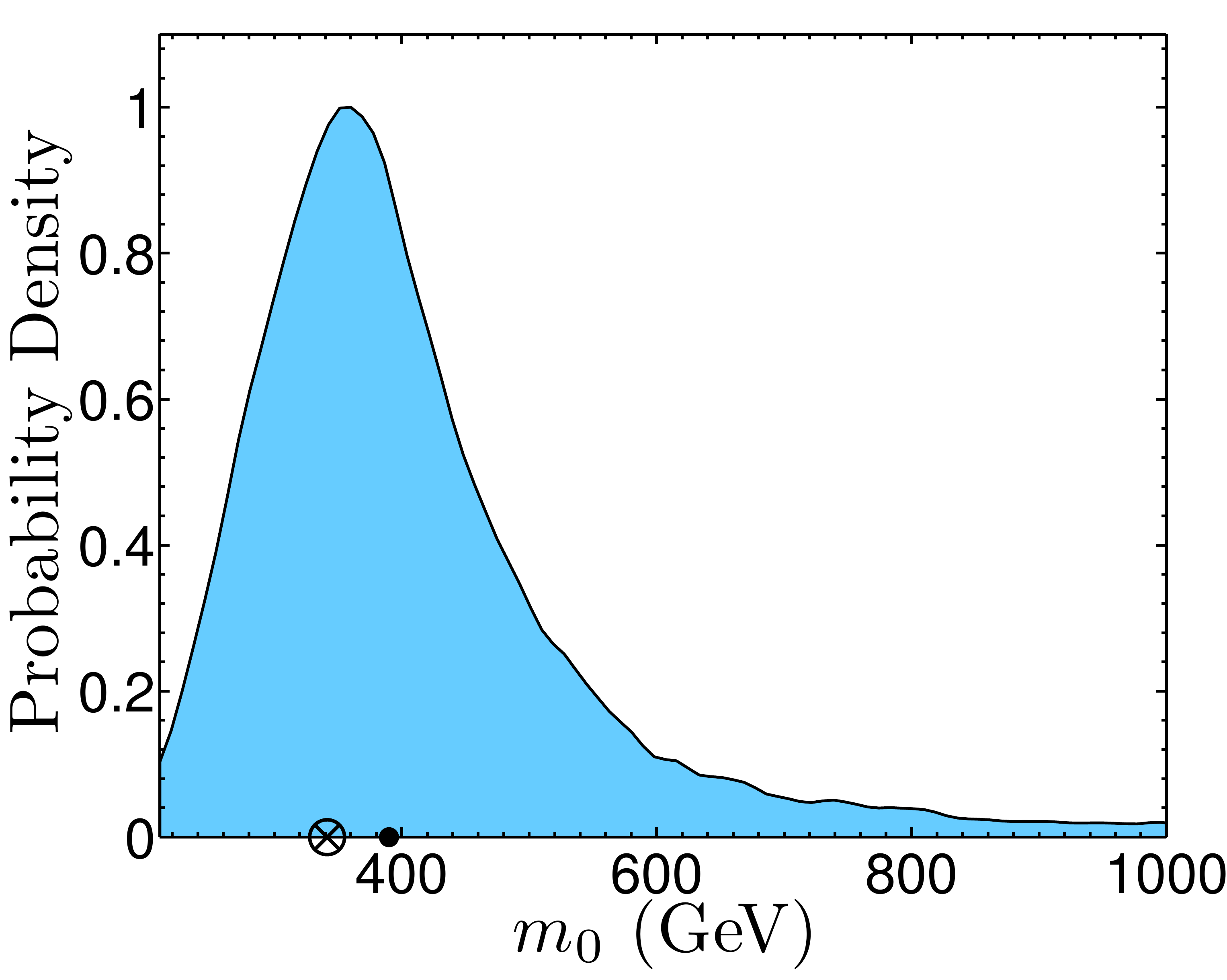}
    \includegraphics[scale=0.14]{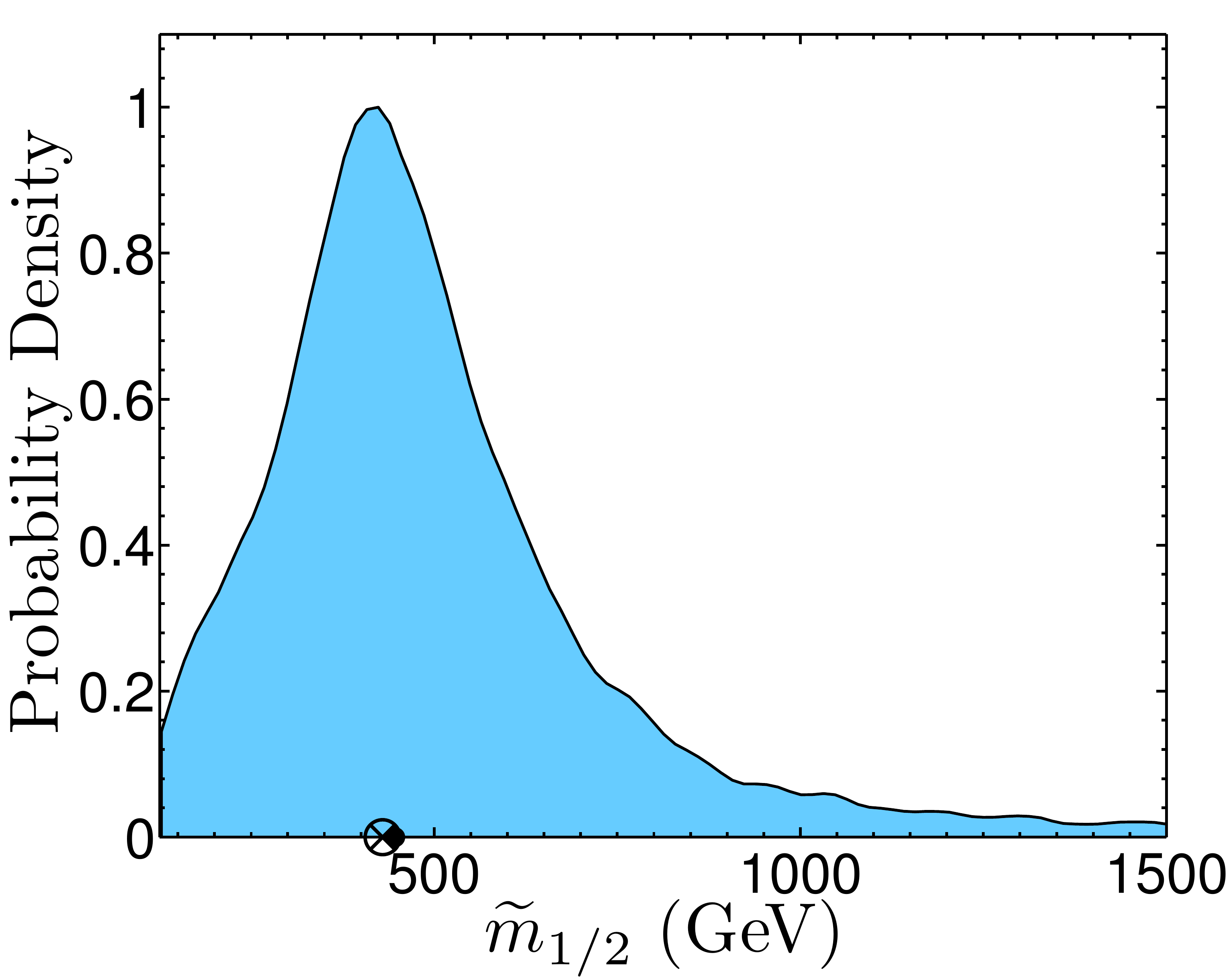}
    \includegraphics[scale=0.14]{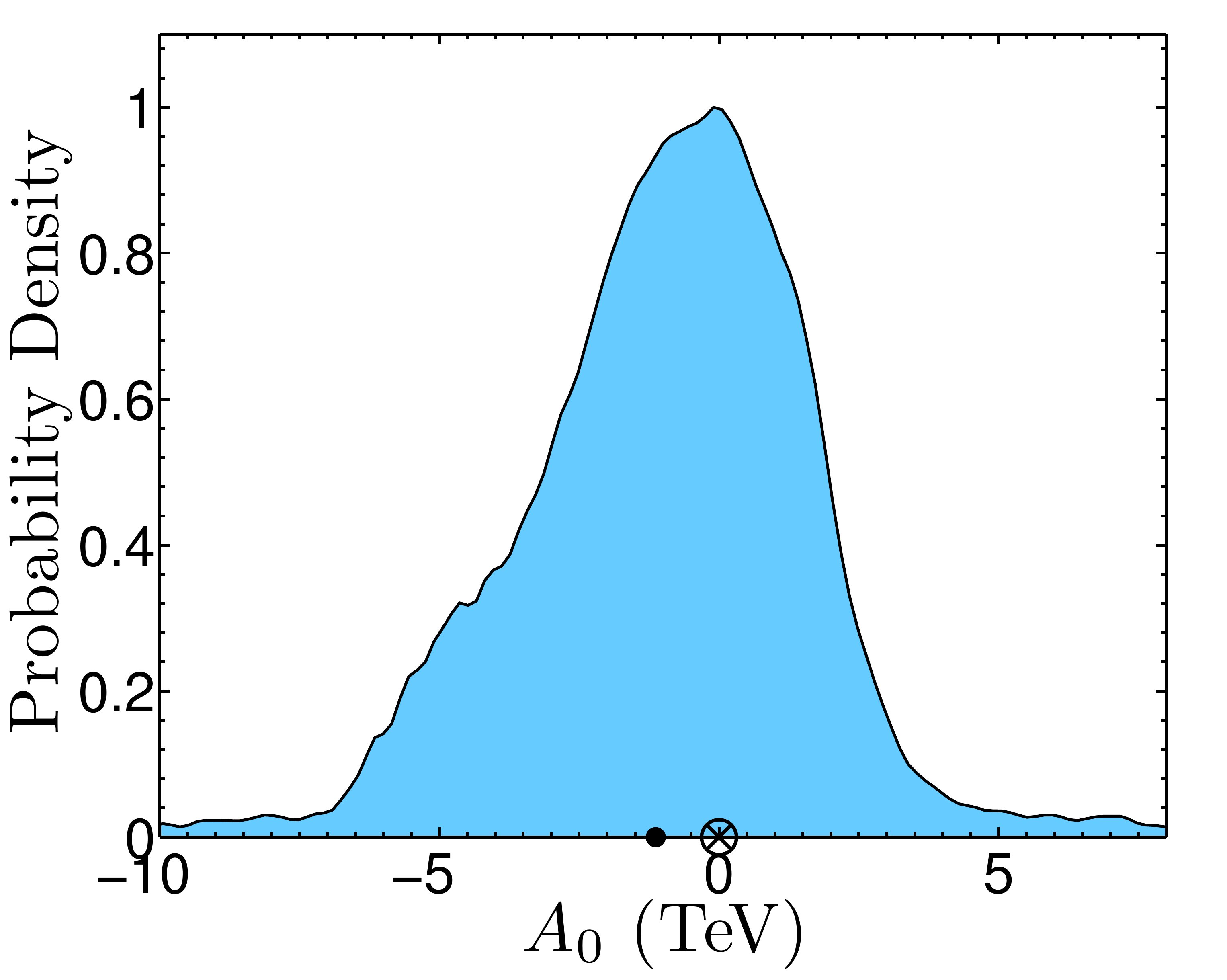}
    \includegraphics[scale=0.14]{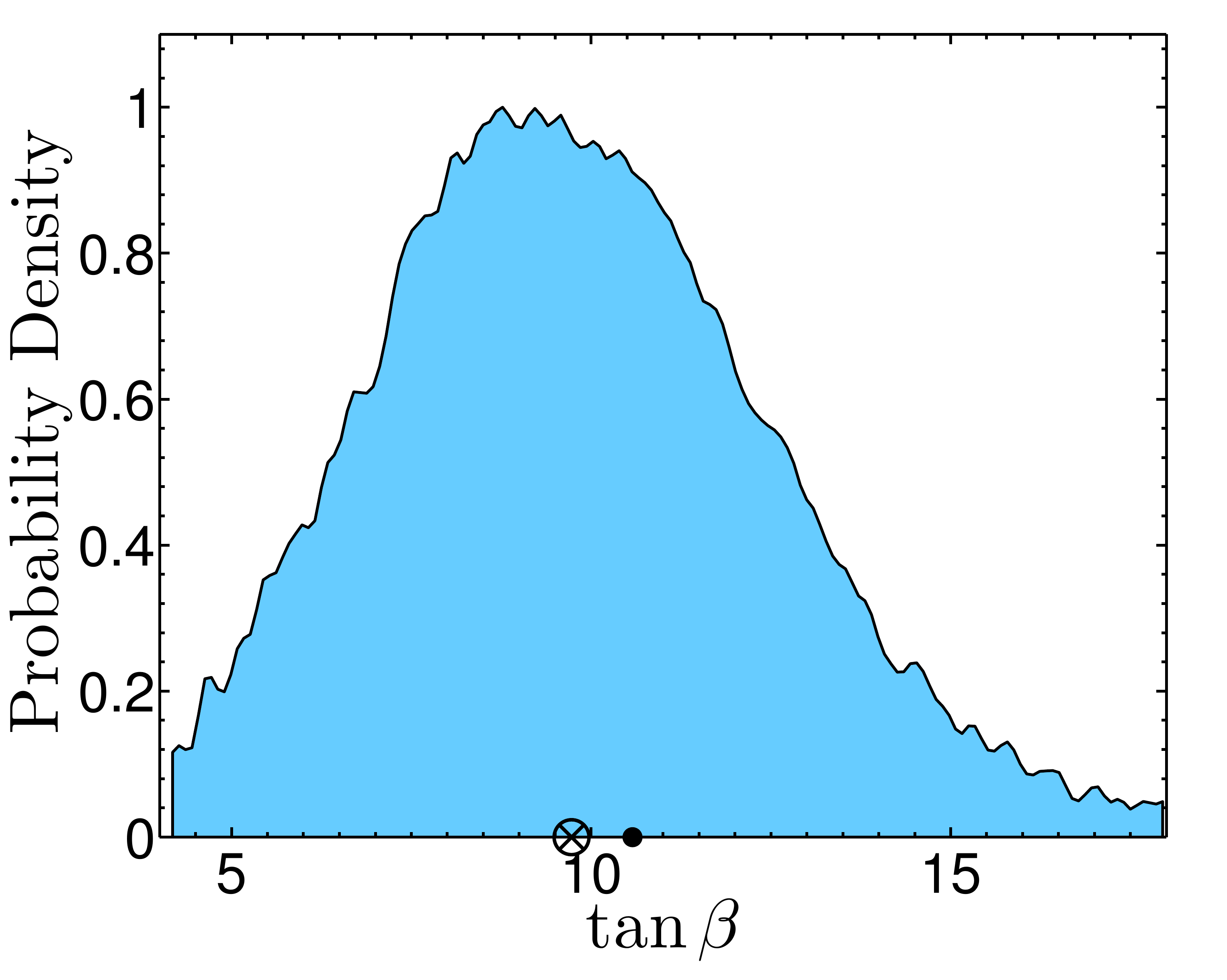} \\
    \includegraphics[scale=0.14]{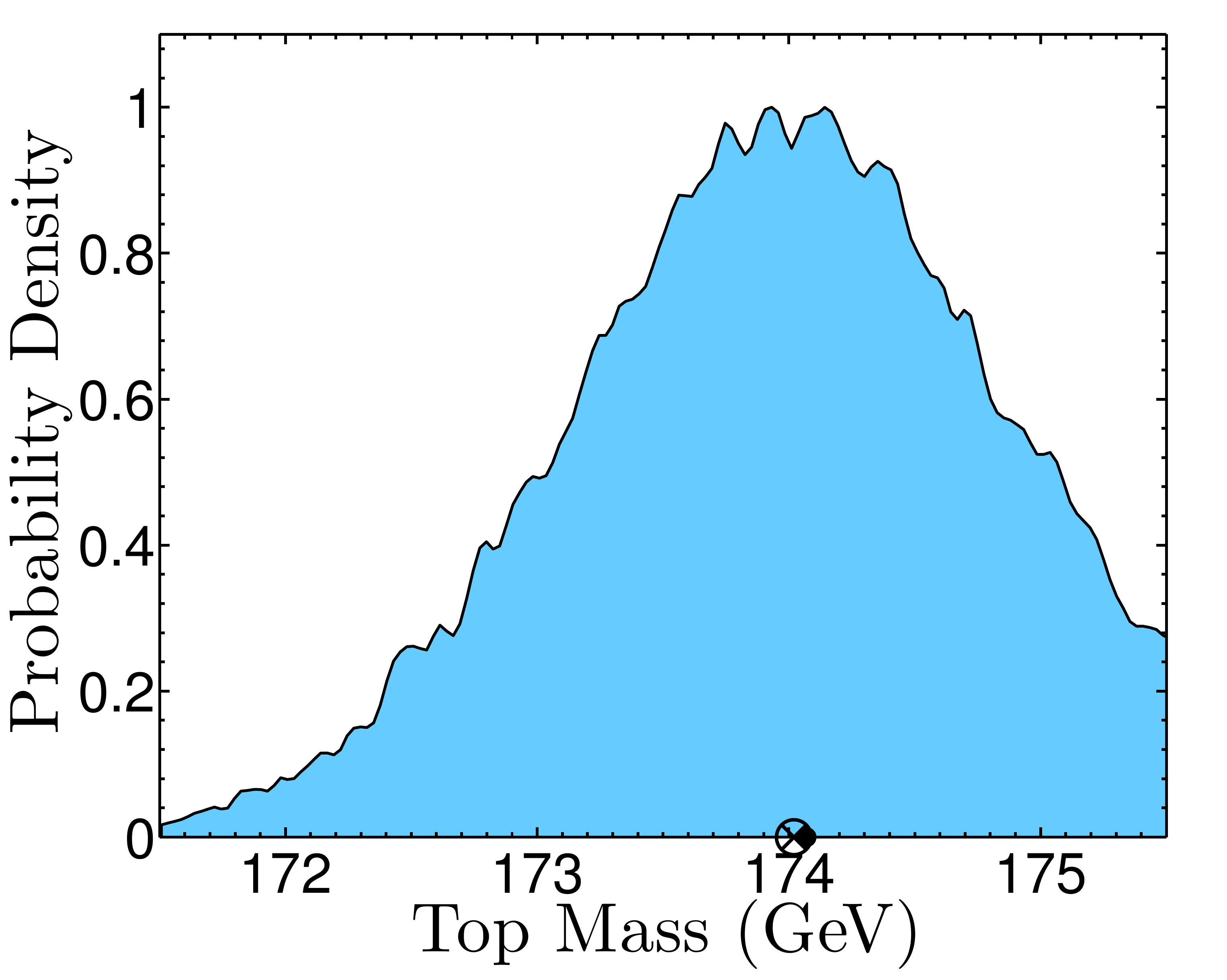}
    \includegraphics[scale=0.14]{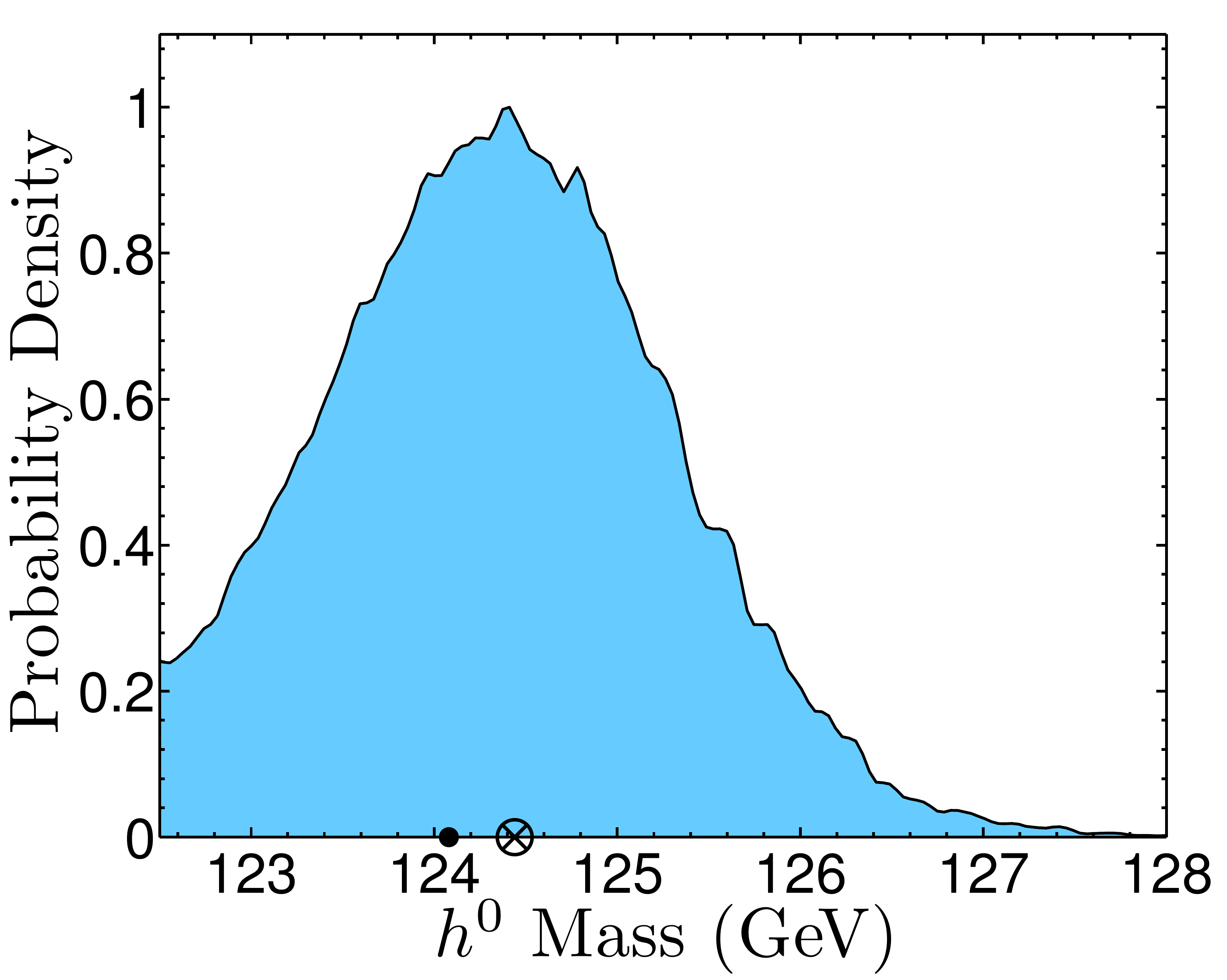}
    \includegraphics[scale=0.14]{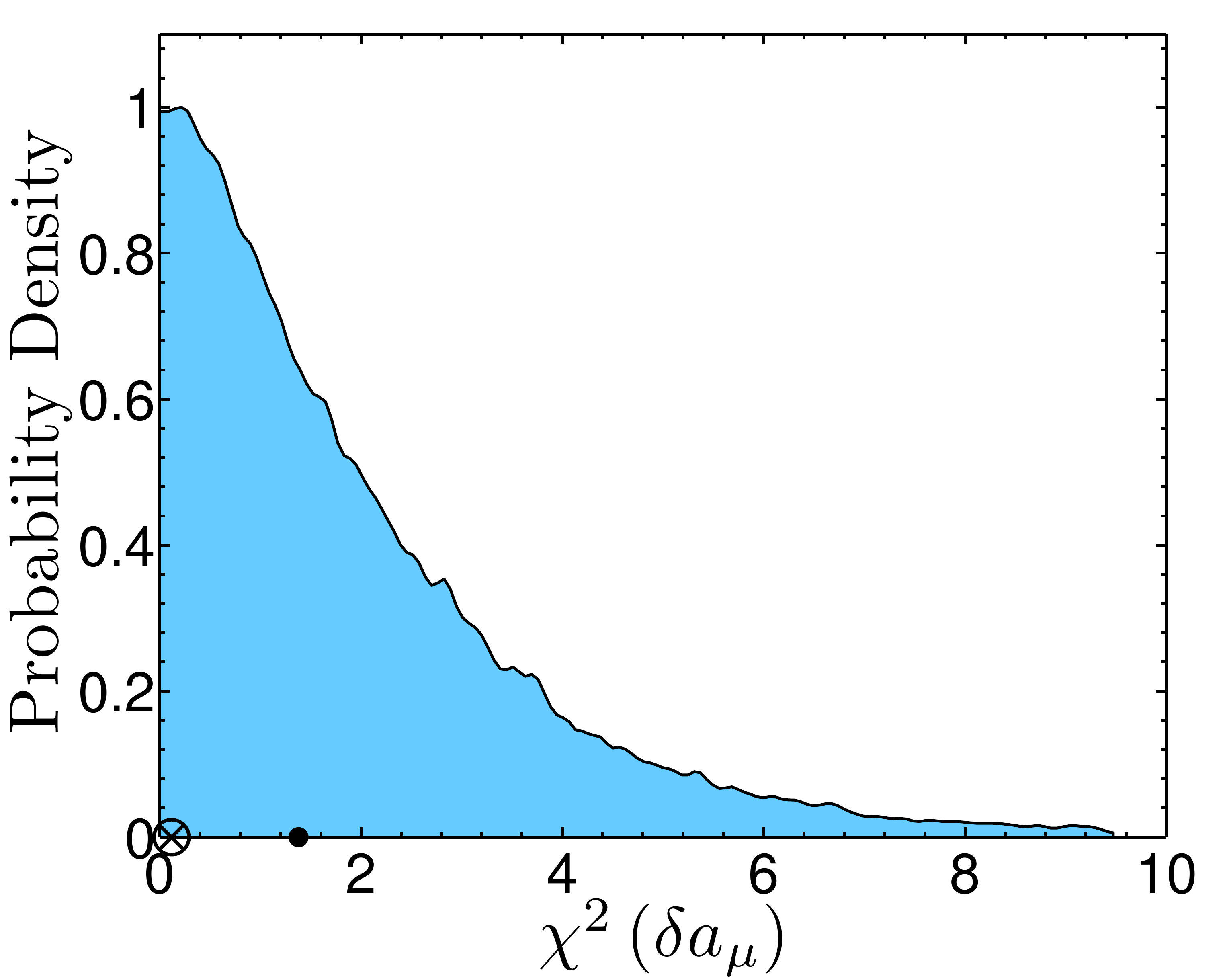}
    \includegraphics[scale=0.14]{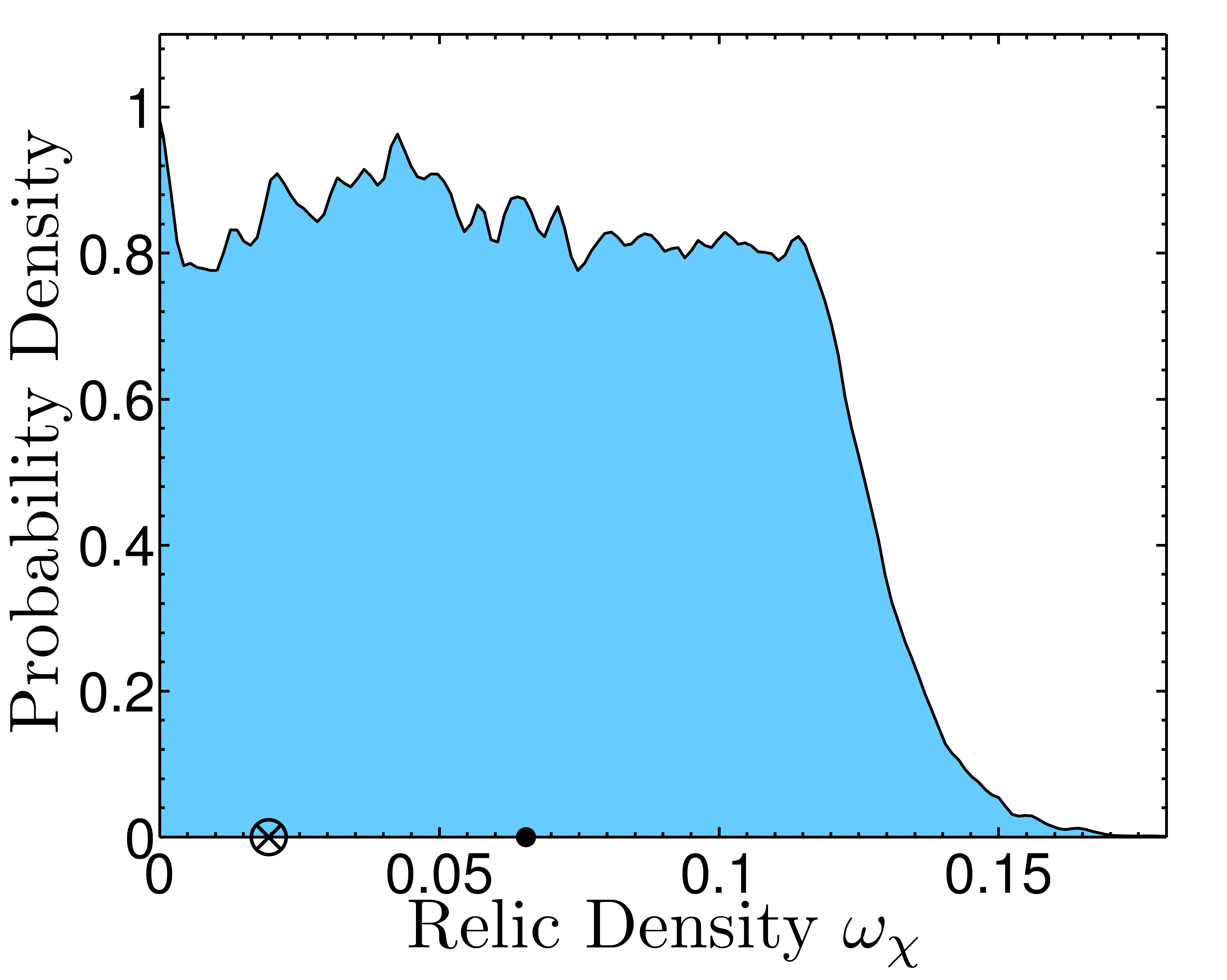}
    \caption{\label{onedplots}
      A display of the marginalized posterior probability distributions for \gsugra in the parameters of interest as well as some 
      important derived quantities. The top row (left to right) gives the posterior PDF for \mo, \mhf, \az, and \tb, 
      and the bottom row (left to right) displays the same for the top quark mass, the light $CP$ even Higgs boson mass, 
      the contribution to $-2\ln\mathfrak{L}$ due to the anomalous magnetic moment of the muon (which we have denoted as 
      $\chi^2(\delta a_\mu)$, and the thermal relic density of cold dark matter, $\omega_\chi$. The location of the best-fit 
      point is indicated by a circled `X' and the posterior mean is given with a solid dot.
    }
  \end{center}
\end{figure*}

Our analysis was performed using our software package \susykit~\cite{susykit}, which uses the efficient multi-modal ellipsoidal nested 
sampling algorithm implemented in the {\sc MultiNest}~\cite{Feroz:2007kg, Feroz:2008xx, Feroz:2011bj} library. Additionally, \susykit interfaces with several standard 
numerical codes such as {\sc SOFTSUSY}~\cite{Allanach:2001kg}, {\sc MicrOMEGAs}~\cite{Belanger:2008sj, Belanger:2010gh}, 
{\sc FeynHiggs}~\cite{Heinemeyer:1998yj, Hahn:2010te}, and {\sc SuperIso Relic}~\cite{Mahmoudi:2009zz, Arbey:2011zz}. \susykit is written 
entirely in C++11 and is largely inspired by the FORTRAN-90 code {\sc SuperBayes}~\cite{deAustri:2006pe, superbayes}.

We specify the {\sc MultiNest} sampling parameters $n_\mathrm{live}=\text{5,000}$ and $\mathtt{tol}=0.01$. The analysis has required 
the evaluation of the likelihood function at 1.1 million points to sufficiently explore the parametric space.
The result is a chain of 81,000 Monte Carlo sample points which is used to compute 1D and 2D marginalized 
distributions in our principal and derived parameters, and to establish credible regions in these parameters. We found 
that the credible regions entered areas that would be excluded by the LHC in minimal SUSY GUT models such as mSUGRA, so we 
found it necessary to evaluate the impact of LHC searches on \gsugra.

\section{LHC Analysis\label{lhc}}
In order to evaluate the impact of null results in the searches for supersymmetry at the LHC on \gsugra, we construct 
an auxiliary likelihood function, $\mathfrak{L}_\mathrm{LHC}$, based on the Monte Carlo event generation and detector simulation 
for our sample points.

We begin by generating 200,000 events for each sample point in our chain using {\sc PYTHIA}~\cite{Sjostrand:2007gs, Sjostrand:2006za} considering $2\to2$ SUSY production processes 
with $\cm=8\TeV$. We find that the total cross section for these processes is  \order{100\fb} and the dominant modes involve the production of \na, \nb, \cha, \slep, \sta, \stb, and, 
\slep and \snu. This is to be expected because in \gsugra, the scalar quark fields all become heavy as they are renormalized to 
the electroweak scale, while the scalar leptons are allowed to remain light to produce contributions to $\delta a_\mu$ and the 
Higgs diphoton decay rate. By investigating the dominant decays of these particles, we decide that supersymmetry searches in leptonic 
final states are the most relevant to \gsugra. We have used the $3\ell$ and same-sign $2\ell$ searches at CMS~\cite{CMS-PAS-SUS-12-022} using 9.2\ifb 
at $\cm=8\TeV$ to construct our $\mathfrak{L}_\mathrm{LHC}$. These searches are performed using 108 and 4 event bins respectively, which serve 
as counting experiments and are naturally Poisson distributed. Therefore $\mathfrak{L}_\mathrm{LHC}$ is computed by 
\begin{equation}
  \mathfrak{L}_\mathrm{LHC} = \prod\limits_{i\in\mathrm{bins}} \mathfrak{L}_i ~.
  \label{lhclike}
\end{equation}

\begin{figure}[t!]
  \begin{center}
    \includegraphics[scale=0.27]{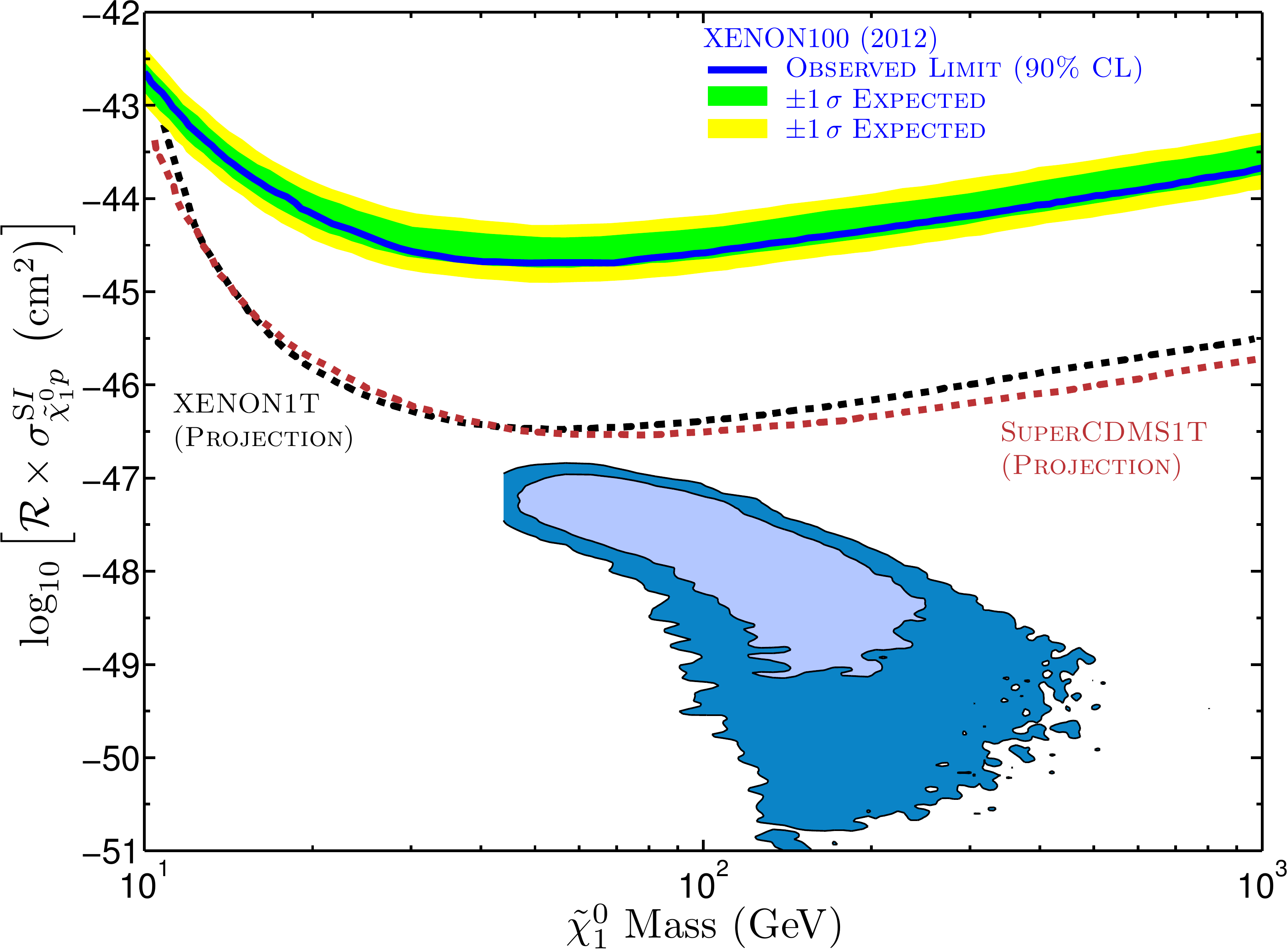}
    \caption{\label{dm}
      A display of the $1\,\sigma$ and $2\,\sigma$ credible regions of the marginalized posterior PDF of \gsugra 
      in the plane of the spin-independent $p$--\na cross section and the \na mass. The current limit from 
      XENON100 is displayed as well as the projected sensitivities for XENON1T and SuperCDMS1T. 
    }
  \end{center}
\end{figure}

Each $\mathfrak{L}_i$ would be a simple Poisson likelihood, except that one of the parameters to the Poisson distribution, the expected background 
yield, $b_i$, can have a large uncertainty, $\delta b_i$. Thus, it is necessary to convolve the Poisson distribution with a distribution 
for the background yield. Na\"ively this would be a Gaussian distribution, however in the case that the relative error in the background 
yield is large, i.e., $\delta b_i / b_i \gtrsim 20\%$, then a non-trivial portion of the convolution is due to contributions from 
negative $b_i$, or even if the integration is limited to non-negative background, a large portion of the PDF may be omitted. 
Thus as a heuristic, 
we use the following definition for \nolinebreak $\mathfrak{L}_i$:
\begin{equation}
  \mathfrak{L}_i = \int_0^\infty \pois(s_i+\bar{b};o_i) F(b_i,\delta b_i; \bar{b}) ~,
  \label{lhclikebin}
\end{equation}
where $i$ is the event bin, $\pois$ is the Poisson probability mass function, $s_i$ is the expected signal yield, $o_i$ is the 
number of observed events, and as defined already $b_i$ is the expected background yield, and $\delta b_i$ is the uncertainty 
in the background. The function F is defined according to our heuristic 
\begin{equation}
  F(b_i,\delta b_i; \bar{b}) =
    \begin{cases}
      \mathcal{N}(b_i,\delta b_i; \bar{b}) \text{ , $\delta b_i/b_i < 20\%$} \\
      \ln\mathcal{N}(b_i,\delta b_i; \bar{b}) \text{ , $\delta b_i/b_i \ge 20\%$}
    \end{cases} ~,
    \label{kernels}
\end{equation}
where $\mathcal{N}$ is the Gaussian distribution and $\ln\mathcal{N}$ is the log-normal distribution.
As a further heuristic, it is necessary to account for cases when either $b_i=0$ or $\delta b_i=0$. These cases are clearly 
oversights in the CMS {\it preliminary} analysis summary; still they must be addressed. We choose a sentinel value $\Delta = 10^{-6}$
and use $\delta b_i = \Delta$ if $\delta b_i$ is zero and we set $b_i = \delta b_i$ if $b_i$ is zero.

The expected signal yield $s_i$ is the product of the efficiency $\epsilon_i$ with the total SUSY cross section and the integrated 
luminosity. The efficiency $\epsilon_i$ is the proportion of the total generated events that would be counted in the $i^\mathrm{th}$ bin, 
and is determined by running the events through a detector simulation, which we have carried out with {\sc PGS4}~\cite{Conway:PGS}. Jet objects were reconstructed 
using the anti-$k_\mathrm{T}$ algorithm, with a distance parameter of 0.5. We implemented the cuts to place events into bins in a modified 
version of {\sc Parvicursor}~\cite{parvicursor}. The object selection criteria, event vetoes, and geometrical cuts are reproduced as in~\cite{CMS-PAS-SUS-12-022}.

\begin{figure}[t!]
  \begin{center}
    \includegraphics[scale=0.30]{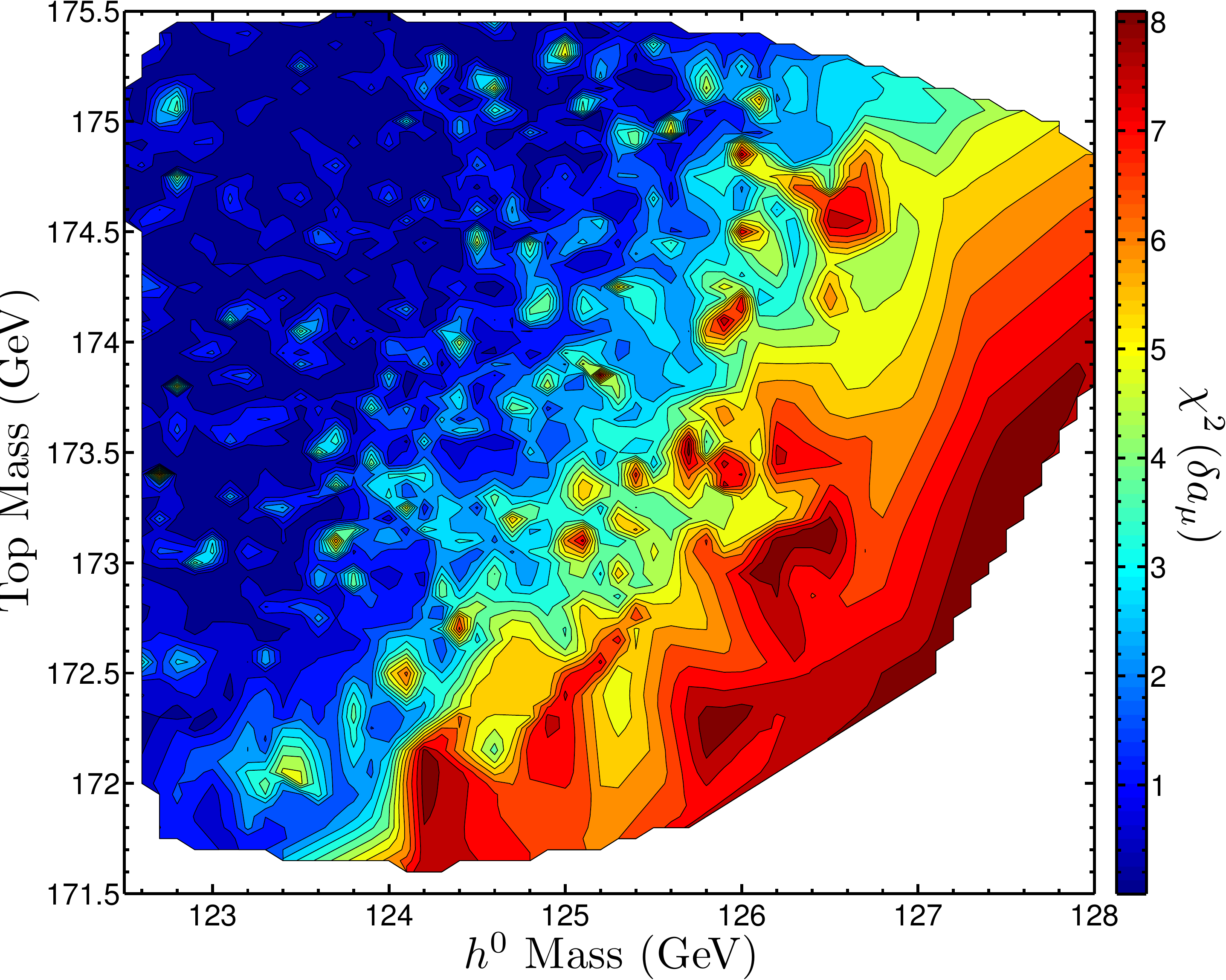}
    \caption{\label{gm2}
      A display of level curves in the statistic $\chi^2\left(\delta a_\mu\right)$, which is the contribution to 
      $-2\ln\mathfrak{L}$ due to $\delta a_\mu$. The level curves are given in the plane of the top mass and $h^0$ mass.
      The level curves are constructed by interpolating equally-weighted sample points.
    }
  \end{center}
\end{figure}

To combine the likelihood from these searches to the likelihood function described in \cref{framework}, we first compute the likelihood for the 
Standard Model according to this analysis by turning off the signal, $\mathfrak{L}_\mathrm{SM} = \left.\mathfrak{L}_\mathrm{LHC}\right|_{s=0}$. 
We then add the likelihood ratio statistic to the full likelihood function,
\begin{equation}
  -2\ln\mathfrak{L} \to -2\ln\mathfrak{L} -2\ln{\left( \min\left\{ \frac{\mathfrak{L_\mathrm{LHC}}}{\mathfrak{L_\mathrm{SM}}}, 1\right\} \right)} ~,
  \label{likecombine}
\end{equation}
which is approximately $\chi^2$ distributed, and is a natural addition to the other ``pull'' terms in our likelihood function.
Having computed the updated likelihood due to these CMS searches, it is necessary to re-weight the samples by a factor $\exp\left(\Delta\ln\mathfrak{L}\right)$. 
We can now proceed to determine the marginalized posterior probability distributions within our parameters of interest.

\section{Results \label{results}}

In this section we present the results from our Bayesian analysis. Given our likelihood function, we determine the Bayesian 
evidence of \gsugra to be \linebreak $\ln\mathcal{Z} = -11.9 \pm 0.042$. We provide this for reference, as we do not perform a model 
selection test. The best-fit point in our analysis is determined to have $\chi^2_\mathrm{min} = 2.73$, and leaving out some of the 
nuisance parameters, is specified by $(\mo,\mhf,\az,\tb,m_t^\mathrm{pole}) = (341,429,298,9.73,174)$ where the massive parameters are specified 
in \gev. This point illustrates the general result of \gsugra that high \hz mass and $\delta a_\mu$ can be simultaneously satisfied. Additionally, 
the large scalar quark and gluino masses allow for consistency with \br\bsmumu and \br\bsg. The credible regions in the masses of the heavier 
particles in \gsugra are presented in the right panel of \cref{masspost}, and the light particles of \gsugra that create the $\delta a_\mu$ 
contribution as well as the contribution to the diphoton Higgs decay are given in the left panel.

The $1\,\sigma$ and $2\,\sigma$ credible regions in our parameters of interest are given in \cref{twodplots}, where we have chosen to use the 
dimensionless parameter $A_0/m_0$. The 1D posterior distributions in these parameters are given in the top panels of \cref{onedplots}, though 
here we did give the distribution for the dimensionful parameter $A_0$.

While \gsugra largely achieves the correct \hz mass and $\delta a_\mu$ contribution as shown in the middle two lower panels of \cref{onedplots}, 
the posterior distribution in the top mass is shifted up from the central value by 0.5\GeV to 174\GeV, which is evident in the lower left panel of 
\cref{onedplots}. The tension between the top mass, the \hz mass and $\delta a_\mu$ is clearly displayed in \cref{gm2} where we have interpolated 
sample points from a slice in our likelihood function and presented level curves in ``$\chi^2(\delta a_\mu)$'' which is the contribution to 
$-2\ln\mathfrak{L}$ due to $\delta a_\mu$. It is evident that the higher \hz mass and $\delta a_\mu$ is best matched in \gsugra for a slightly 
heavier top quark. 

We point out that this tension is not overly significant in \gsugra for two reasons. First, there is a large theoretical uncertainty in the 
calculation of the \hz mass at the 2-loop level, which when considered does lift most of the tension. Next, we specified in \gsugra 
$M_3=10\,\mhf$, where 10 is an arbitrary choice. Allowing the coefficient to be a new degree of freedom or simply selecting several different 
choices will likely resolve this tension as well.

In our Bayesian analysis, we have sampled the parameter space using the older WMAP7 value for $\omega_\chi$ in $\mathcal{L}$ but we can see 
from the fourth panel from the left in the bottom row of \cref{onedplots} that the slightly larger value indicated by WMAP9 and Planck would 
simply enlarge our credible region. Additionally, we see in \cref{dm} that \gsugra is not currently constrained by the best available limit 
on the direct detection of \na dark matter, and is slightly beyond the projected sensitivity of XENON1T and SuperCDMS1T, creating a sort of 
nightmare scenario for dark matter experiments, as our dark matter signal would be competing with the cosmic neutrino background. The LSP in our 
model is consistently a bino, and the \nb is a wino. There is virtually no mixing with the Higgsino sector as the Higgsino mass parameter $\mu$ 
becomes very large due to the large $M_3$.  The sensitivity to dark matter experiments can be increased by adjusting the ratio of $M_1$ to $M_2$ 
to allow for greater bino-wino mixing within the LSP state.

One of the exceptional aspects of \gsugra is the presence of many light superpartners that have thus far evaded detection at the LHC. We concede 
that the searches that we considered here are not by any means comprehensive, but they are designed to constrain the production modes 
most prevalent in \gsugra. The limits are evaded largely due to the stringent selection criteria and the difficulty in identifying $\tau$ leptons. 
Additionally, the mass hierarchy of \gsugra limits the possibility of cascading decays.

We note that the parametric space of \gsugra, naturally fits into the Hyperbolic Branch~\cite{Chan:1997bi, Chattopadhyay:2003xi, Baer:2003wx} of 
radiative breaking of the electroweak symmetry. This is due to the fact that the stop masses are 
driven to be large by the gluino, giving a large $Q=\sqrt{m_{\stopa}m_{\stopb}}$, and it was shown in~\cite{Akula:2011jx, Liu:2013ula}  that $Q\gtrsim1\TeV$ corresponds to a 
hyperbolic geometry of soft parameters that give radiative EWSB (a large SUSY scale in the tens of TeV also arises in 
a certain class of string motivated models~\cite{Acharya:2008zi,Feldman:2011ud}). 
Still, \gsugra as it stands produces a large value of $\mu$ with respect to the $Z$ mass. Specifically, a 
large value of $\mu$ is necessary to balance the large value of $M_3$ which enters in the corrections to the $H_2$ field mass. 

\subsection{Higgs Diphoton Decay \label{diphoton}}
In the Standard Model, the loop-induced decay of the Higgs into two photons is mediated mainly by the $W$, top, and to a lesser extent, the bottom 
quark. The partial width reads~\cite{Djouadi:2005gi}
\begin{equation}
  \Gamma\left(H\to\gamma\gamma\right) = \frac{\alpha^2_\mathrm{EM}m_H^2}{256v^2\pi^3} \left\lvert \sum\limits_{f=t,b} N_{c,f}Q_f^2A_{1/2}(\tau_f) + A_1(\tau_W) \right\rvert^2
  \label{SMHGamGam}
\end{equation}
where $\tau_i$ = $4m_i^2/m_H^2$, and the spin form factors are 
\begin{gather}
  A_{1/2}(\tau) = 2\tau\left(1-(\tau-1)f(\tau)\right) \\
  A_{1}(\tau) = -\left(2 + 3\tau - 3\tau(\tau-2)f(\tau)\right)
\end{gather}
and the universal scaling function $f(\tau)$ is
\begin{equation}
  f(\tau) = \begin{cases}
    \begin{array}{lr}
      \arcsin^2\left(\tau^{-1/2}\right) & : \tau \ge 1 \\
      -\dfrac{1}{4}\left(\ln\dfrac{1+\sqrt{1-\tau}}{1-\sqrt{1-\tau}} - \imath\pi \right)^2 & : \tau < 1
    \end{array}
  \end{cases} ~.
\end{equation}
Supersymmetry corrects this partial width~\cite{Djouadi:2005gj} by factors involving the Higgs mixing angle $\alpha$ and $\beta$ arising from the two Higgs doublets. Additionally, new 
amplitudes are available mediated by the charged Higgs, charginos, and sfermions. The couplings to the charginos arise from Higgsino--gaugino 
mixing, but in \gsugra the Higgsinos are very heavy thus the lighter chargino is always purely charged wino while the heavier
one
 is purely
charged Higgsino. This means that overall the chargino contribution is small either because the coupling is suppressed or because the mass is 
too large. The charged Higgs exchange is also suppressed due to its large mass. Thus the largest contributions can come
only from the sfermion 
sector, which in \gsugra is dominated by the staus.

\begin{figure}[t!]
  \begin{center}
    \includegraphics[scale=0.30]{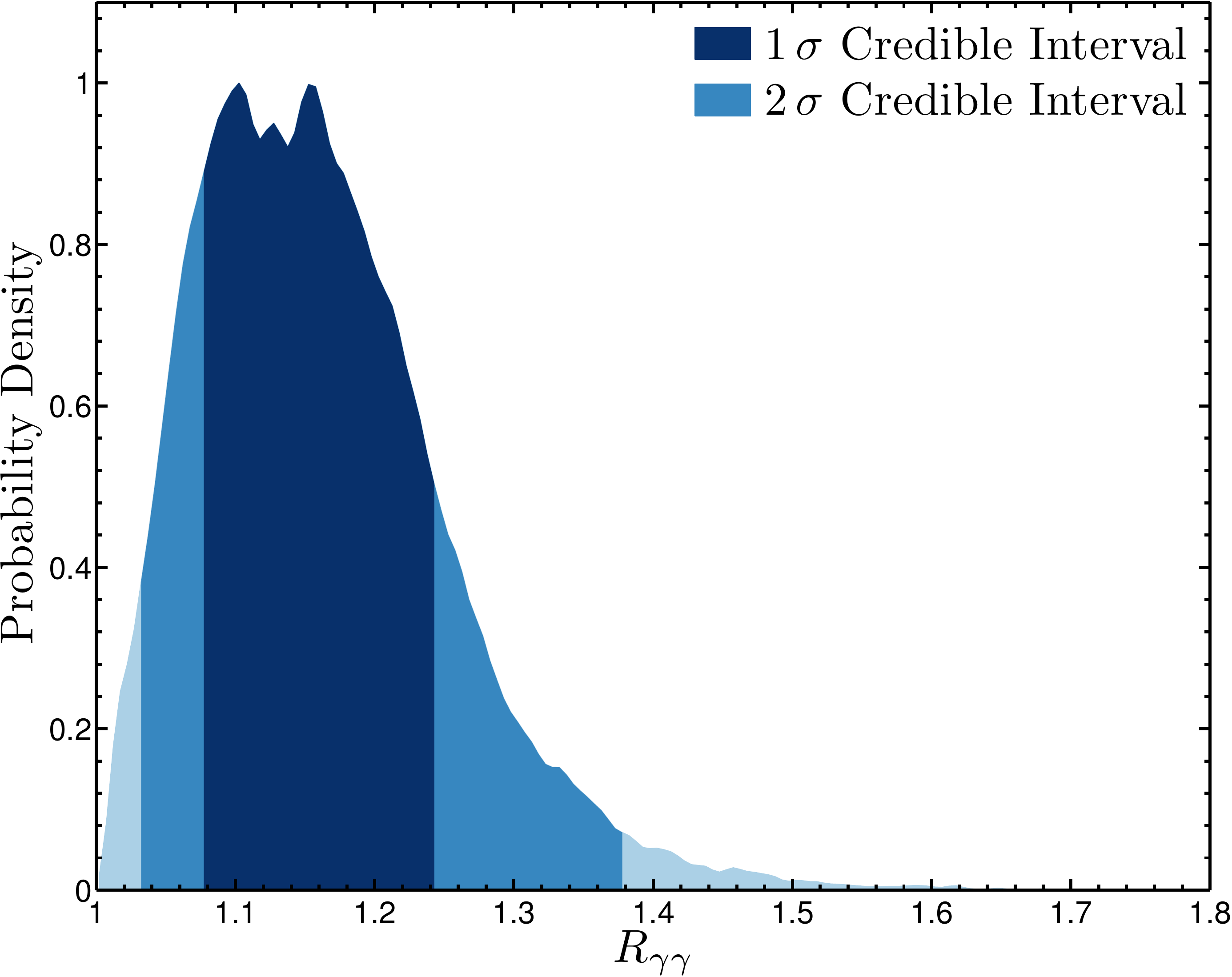}
    \caption{\label{rgg}
    A display of the
      marginalized posterior probability density of $R_{\gamma\gamma}$ from our analysis. The $1\,\sigma$ and $2\,\sigma$ credible intervals are 
      indicated in darker blues. We define $R_{\gamma\gamma}$ as the ratio of the diphoton partial width of the light $CP$-even Higgs boson to the 
      corresponding width for a Standard Model Higgs of the same mass (see \cref{rggdef}).
    }
  \end{center}
\end{figure}

In the decoupling limit where $M_A \gg M_Z$ which corresponds to $\alpha=\beta-\pi/2$, the Higgs coupling to the staus is given by~\cite{Giudice:2012pf, Feng:2013mea}
\begin{multline}
  g_{\hz\stau_i\stau_i} = I_3^\tau c_i \mp Q_\tau\sin^2\theta_\mathrm{W} \cos{2\theta_{\stau}} \\ 
    \mp \frac{m_\tau(A_\ell - \mu\tan\beta)}{2M_Z^2}\sin{2\theta_{\stau}} -\frac{m_\tau^2}{M_Z^2}
\end{multline}
with $c_1 = \cos^2{\theta_{\stau}}$, and  $c_2 = \sin^2{\theta_{\stau}}$. 
The `$-$' case corresponds to $i=1$, and the `$+$' case corresponds to $i=2$.
The partial width in \gsugra including the amplitude due to staus then reads
\begin{multline}
  \Gamma\left(\hz\to\gamma\gamma\right) = \frac{\alpha^2_\mathrm{EM}m_H^2}{256v^2\pi^3} \left\lvert \sum\limits_{f=t,b} N_{c,f}Q_f^2A_{1/2}(\tau_f)\right. \\
     \left. + A_1(\tau_W) + \sum\limits_{i=1,2}g_{\hz\sta\stb}\frac{M_Z^2}{\mstau^2}A_0(\tau_i)\right\rvert^2
\end{multline}
and the spin zero form factor is
\begin{equation}
  A_{0}(\tau) = -\tau(1-\tau f(\tau)) ~.
\end{equation}
We identify the ratio of this partial width to the Standard Model width given in \cref{SMHGamGam} as $R_{\gamma\gamma}$. (We have taken the ratio of the 
theoretical and observed \hz production to be unity.) We compute this ratio for each 
of our Monte Carlo samples and construct the 1D posterior PDF in this derived parameter which we present in \cref{rgg}. We find that \gsugra 
generically produces a $\sim20\%$ boost to this decay mode over the Standard Model case. The $2\,\sigma$ credible interval is $[1.03, 1.38]$, which is 
quite consistent with the preliminary results arriving from the LHC.

\section{Conclusion \label{conclusion}}
The recent observation of the Higgs boson mass around 125\GeV points to large loop corrections which can be achieved with a large weak scale of 
SUSY. A large SUSY scale also explains the suppression of SUSY contributions to the decay \bsmumu, to be consistent with the recently measured  branching ratio for this process. On the other hand, the experimental observation of a $3\sigma$
effect in $\delta a_\mu$  and a possible excess in the diphoton rate $R_{\gamma\gamma}$ in the Higgs boson decay over the standard model prediction
cannot be explained with a high SUSY scale. Thus the two sets of data point to a two scale SUSY spectrum,
one a  high scale consisting of  colored particles, i.e., the squarks and the gluinos,  and the Higgs bosons (aside from the lightest Higgs) and 
the other a low scale  for masses of uncolored particles including sleptons and the electroweak gauginos.

In this work we discuss the high scale supergravity grand unified model, \gsugra, which includes the feature
of a two scale sparticle spectrum where the sparticle spectrum is widely split at the electroweak scale. 
This is accomplished within supergravity grand unification  with non-universal gaugino masses such that
$M_3\gg M_1,M_2,m_0$. As an illustration we consider the specific case where  $M_1:M_2:M_3 = 1:1:10$ at the unification scale,
$M_1=M_2= \tilde m_{1/2}$ and $M_3>>m_0$. This case is designed to be mainly 
illustrative and can be easily embedded within $\SU(5)$ and $\SO(10)$. Using a Bayesian Monte Carlo analysis, It is found that this 
construction simultaneously explains the high \hz mass, null results for squarks and gluino searches at the LHC,
a negligible correction to the branching ratio for \bsmumu, a $3 \sigma$ 
deviation of $g_\mu-2$ from the Standard Model prediction as well as the nascent excess in the diphoton signal 
of the Higgs.

The observable sparticle spectrum at the LHC in this model consists of  light sleptons and light electroweak gauginos.
However, sleptons and electroweak gauginos are typically difficult to observer at the LHC and thus far have
evaded detection in multi-lepton searches in experiments at the ATLAS and the CMS detectors with the
7 \TeV and 8 \TeV data. The most promising $2\to 2$ processes that can generate sparticles at the LHC in this
model are $pp\to \cha\chabar,\nb\cha$. The identifying signatures of such processes will indeed be 
multi-leptons and missing energy.  It is hoped that at increased energies and with larger luminosities 
such signals will lie in the observable region. However, a detailed analysis of the signals in needed 
requiring a knowledge of the backgrounds for these processes. 

Another aspect of the simplified \gsugra model relates to the spin-independent 
$\tilde \chi_1^0-p$ cross section.  This cross section is found to be rather small for the case when the 
gaugino masses are chosen in the ratio $(1:1:10)$. The reason for this smallness is easily understood.
The constraint $M_1 = M_2$ at the GUT scale, leads to an LSP which is essentially purely  bino
with very little Higgsino or wino content.  The purely bino nature of the  LSP leads to a suppressed
$\tilde \chi_1^0-p$ cross section (see e.g.,~\cite{Feldman:2010ke}) which lies beyond the reach of the current 
and projected sensitivities for direct-detection experiments. However,  the above result is very specific
to the $M_1:M_2:M_3=1:1:10$ assumption and a modification of the above should allow 
 $\tilde \chi_1^0-p$ cross section within the observable range in the projected  sensitivities for direct-detection experiments.
We note that while our analysis was performed using the older WMAP7 measurement of the cold dark matter relic density, the newer measurements 
from WMAP9 and Planck (with 15.5 months of data) only slightly increase the measurement. As we only apply the upper limit from these measurements 
to allow for the possibility of multi-component theories of dark matter, the newer results would only expand the credible regions of our parameter 
space and either increase or not affect at all the likelihood of our best-fit point.

Finally, we  note that the large squark masses in \gsugra would also help stabilize the proton against decay from baryon and lepton number violating dimension
five operators~\cite{Arnowitt:1993pd,Liu:2013ula,Hisano:2013exa} (for a review see~\cite{Nath:2006ut}). 

\begin{acknowledgments}
One of us (S.A.) thanks Darien Wood for helpful discussion of the methodology used in \cref{lhc}.
This research is  supported in part by NSF grants PHY-0757959 and PHY-0969739,  
and by XSEDE grant TG-PHY110015. This research used resources of the National Energy Research Scientific Computing 
Center, which is supported by the Office of Science of the U.S. Department of Energy under Contract No. DE-AC02-05CH11231 
\end{acknowledgments}

\bibliography{master}

\end{document}